\documentclass[onecolumn]{mn2e}
\usepackage{graphics}
\input{epsf}
\def\bvarphi{\mbox{\boldmath $\varphi$}}
\def\bphi{\mbox{\boldmath $\phi$}}
\def\bsigma{\mbox{\boldmath $\sigma$}}
\newcommand{\beq}{\begin{equation}}
\newcommand{\eeq}{\end{equation}}
\def\la{\hbox{\raise.35ex\rlap{$<$}\lower.6ex\hbox{$\sim$}\ }}
\def\ga{\hbox{\raise.35ex\rlap{$>$}\lower.6ex\hbox{$\sim$}\ }}

\def\beq{\begin{equation}}
\def\eeq{\end{equation}}
\def\beqa{\begin{eqnarray}}
\def\eeqa{\end{eqnarray}}

\def\order#1{{\cal O}\left({#1}\right)}
\newcommand{\sfrac}[2]{ \mbox{$\frac{#1}{#2}$}}
\title[Hydrodynamical stability of thin accretion discs]{Hydrodynamical stability of thin accretion discs:
transient growth of global axisymmetric perturbations\thanks{Research supported by
    the Israel Science Foundation, the Helen and Robert Asher
    Fund and the Technion Fund for the Promotion of Research}}
\author[O.M. Umurhan et al.]{O.M. Umurhan$^{1,3}$\thanks{Email: mumurhan@physics.technion.ac.il},
A. Nemirovsky$^{1}$, O. Regev$^{2,4,5}$ and
G. Shaviv$^{2}$\\
$^{1}$Department of Physics, Technion-Israel Institute of
Technology, 32000 Haifa, Israel\\
$^{2}$Department of Physics and the Asher Space Research Institute, Technion-Israel Institute of
Technology, 32000 Haifa, Israel\\
$^{3}$Department of Astronomy, City College of San Francisco, San Francisco, CA 94112, USA \\
$^{4}$Department of Astrophysics, American Museum of Natural History, New York, NY 10024, USA\\
$^{5}$Department of Astronomy, Columbia University, New York, NY 10028, USA}

\date{Accepted ------------. Received ------------------}

\pagerange{\pageref{firstpage}--\pageref{lastpage}} \pubyear{2005}

\begin{document}

\label{firstpage}

\maketitle

\begin{abstract}
The purpose of this paper is to explore how accretion discs manifest the
phenomenon of transient growth on a global scale.
We investigate analytically the time response of a thin accretion disc to
particular  axisymmetric perturbations.
To facilitate an analytical treatment we replace the energy equation
with a general polytropic assumption.
The asymptotic expansion of Klu\'zniak \& Kita (2000), which extended
the method of Regev (1983) to a full steady polytropic disc (with $n=3/2$),
is further developed and implemented
for both the steady (for any polytropic index)
and time-dependent problems.
The spatial form and temporal behaviour of
selected dynamical disturbances are studied in detail.
We identify the perturbation space which leads to transient
growth and provide analytical solutions
which manifest this expected transient growth behaviour.  Three terms (physical causes)
responsible for the appearance of transient growth are identified. Two depend
explicitly on the viscosity while the third one is relevant also for inviscid discs.
The main conclusion we draw is that the phenomenon of
transient growth exists in  discs on a global scale.
\end{abstract}

\begin{keywords}
hydrodynamics -- accretion discs
\end{keywords}

\section{Introduction}
\label{introduction}

It is widely accepted that accretion discs form whenever
a sufficiently cool gas,
endowed with a significant amount of angular momentum, is
gravitationally attracted by a relatively compact object. This situation
is quite common in astrophysics and, thus, the study, both observational
and theoretical, of accretion discs has been quite intensive
(see e.g. Pringle, 1981; Frank, King \& Raine, 2002;
for an overview).

When accretion discs were theoretically proposed
(Shakura \& Sunyaev, 1973; Lynden-Bell \& Pringle, 1974)
it had been recognized at the outset
that the angular momentum transport necessary for accretion cannot
take place unless some kind of enhanced effective transport
process is invoked (typical values of the microscopic viscosity
coefficient are by far too small to explain the observationally
inferred accretion rates). As is well known, fluid turbulence induces
anomalous transport and therefore angular momentum transport
with the help of turbulent eddy viscosity has been
proposed to operate in accretion discs. A detailed theoretical
understanding of turbulence is still lacking and,
therefore, the effective viscosity in discs has usually
been approached in a phenomenological way: by
parameterizing the effective viscous coefficient on the basis of
dimensional arguments. This simple approach has been
exceptionally fruitful, giving rise to successful interpretations
of observational results (see Lin \& Papaloizou, 1996;
Frank, King \& Raine, 2002; for reviews).

The `common knowledge' that turbulent flows arise by a transition
from laminar flows, which become {\em linearly} unstable
at a particular critical Reynolds number ${\rm Re_{c}}$, has largely been
based on a number of well-studied experimental and theoretical
flows.  However by the early 1990's the consensus within the astrophysical
community was that this assertion appeared not to be
applicable to thin Keplerian accretion discs (which are relevant to many
astrophysical applications, in which the gas in the disc cools
efficiently), because no definite linear {\em hydrodynamical}
instabilities had been identified in such basic flows. Moreover,
according to the famous Rayleigh circulation criterion, Keplerian
shear flows are expected to be linearly stable.
The magneto-rotational instablity, which has been shown
by Balbus \& Hawley (1991) to operate
in such flows (when the fluid is electrically conducting and for
not too large initial magnetic fields),
provides a linear instability mechanism for accretion discs
and has since become accepted as the driver for
magnetohydrodynamical turbulence and the resulting angular
momentum transport.
It has since become the operating paradigm in the astrophysical
community that purely hydrodynamical turbulence in Keplerian discs
is altogether ruled out - this conclusion is based largely on
the paucity of hydrodynamic turbulent activity in
numerical simulations (see Balbus, 2003; for a review and
references).

The question whether {\em hydrodynamical} turbulence can occur
in accretion discs has however recently been re-opened in the
astrophysical literature, close to a decade after
Balbus , Hawley \& Stone (1996)
appeared to close this matter once and for all.
The reason for this renewed interest in hydrodynamical stability
of accretion discs appears to be three-fold. On the one hand,
some instances of astrophysical systems have emerged in which
the efficiency of the magnetorotational instability may be
questioned (see e.g. Menou, 2000; Sano {\em et} al., 2000;
Fromang, Terquem \& Balbus, 2002). Secondly, a number
of authors have applied to the accretion disc
problem some recent ideas, familiar
in the fluid-dynamical community, about the onset of turbulence
and sustained dynamical activity in shear flows via the mechanism
of transient growth (see below).  Lastly, linear instabilities of
shearing flows appear where gravity plays a catalyzing r\^ole.  These
{\it stratorotational instabilities} (Dubrulle, 2004), which are distinct from
instabilities giving rise to buoyant convection, have been identified
from linear stability analysis (Molemaker et al. 2001, Yavneh et al. 2001,
Dubrulle et al. 2004
Shalybkov \& Rudiger, 2005) and have been proposed to operate
in Keplerian discs.  Yavneh et al (2001) demonstrate that this effect
leads to significant non-linear activity in small-gap limit simulations of
Taylor-Couette flows in a uniform gravitational field.  It seems also possible
that this instability may be responsible for the appearance of long-lived
vortices in the localized hydrodynamical simulations of 3D Keplerian flows of
Barranco \& Marcus (2004).  Though these are interesting, necessary and important
avenues of exploration, these matters are not the subject of our attention
in this work.

The concept of {\em transient growth} (TG for short) of linearly stable modes and
its possible role in giving rise to a {\em bypass} (of linear instability)
transition to turbulence has been  applied extensively to laboratory
shear flows, which appear to become turbulent at Reynolds numbers
that are significantly lower than ${\rm Re_c}$ (given by the usual
modal approach of linear stability analysis).
In particular, for some archetypical flows
(plane Couette, pipe Poiseulle), the theoretical critical Reynolds number
is infinite, that is, these flows are found to be linearly stable.
Yet such flow become clearly turbulent in the laboratory.
The central idea is based on the fact the relevant operator,
arising in linear stability analysis of shear flows, is generally
non-normal (not commuting with its adjoint). Thus the usual procedure
of finding the eigenvalues by solving the appropriate {\em boundary
value} problem, may miss some important features of the linear problem.
Even if all the eigenvalues imply stability (i.e.
have negative or zero real parts), some initial
perturbations may exhibit transient growth, because the {\em eigenvectors}
of a non-normal operator may be non-orthogonal. The correct way
to uncover such behaviour must then rely on solving the
corresponding {\em initial value} problem.
To save space, we shall refer the reader to a recent book
(Schmid \& Henningson, 2001) and a review paper (Grossmann, 2000),
in which the idea that TG may trigger
a transition to turbulence in shear flows is explained in detail,
together with a comprehensive list of references. The
astrophysical papers which will be mentioned shortly contain
also, in their Introduction
sections, short reviews on the history of this idea and its
application to astrophysics.

Among the papers in the astrophysical literature Ioannaou \& Kakouris (2001)
were the first to apply these concepts to the stability of accretion
discs. They studied the {\em global} behaviour of  incompressible two-dimensional
(in the disc plane) perturbations, found copious TG
for optimal perturbations and suggested that a turbulent state
may be maintained by continuous random forcing. Most of the subsequent
works utilized a local {\em shearing box} approximation\footnote{see
Appendix A of Umurhan \& Regev (2004) for a systematic discussion
of this approximation}
and found linear TG in two dimensions (Chagelishvili et al., 2003)
and in even in three dimensions (Tevzadze et al. 2003, Yecko, 2004). It was found
that the third dimension may limit the very large amplification
factors, but if the vertical scale is not too large the growth
is still substantial. Yecko (2004) analyzed the linear three-dimensional
problem and found that strong rotation essentially
`two-dimensionalizes' the most amplified transient disturbances
whose maximal magnitude, for Keplerian shear, scale as ${\rm Re}^{2/3}$.
In two very recent contributions Mukhopadhyayi, Ashfordi \& Narayan (2005)
and Ashfordi, Mukhopadhyayi \& Narayan (2005) conducted detailed
studies of the linear problem in various conditions and supplied
estimates of the Reynolds number needed to sustain TG induced turbulence
in accretion discs. They found that values
of at least  $10^5-10^6$ are needed to overcome the stabilising
influence of Keplerian shear (i.e. Rayleigh stable) and three-dimensionality.

Recently, Umurhan \& Shaviv (2005) employed a global asymptotic approach
(utilizing the natural small parameter of the problem
$\epsilon = \tilde H/\tilde R$,
that is, the thickness of the disc relative to its typical radial scale)
in the study of the dynamical evolution of global axisymmetric disturbances
in an {\em inviscid} 3D disc.
Such an asymptotic approach was introduced for the first time
to the study of thin viscous accretion discs by Regev (1983)
in the context of accretion disc boundary layers. It was later
further developed and used in a remarkable analytical work by
Klu\'zniak \& Kita (2000) to solve for the {\em steady} structure
of a polytropic viscous (using the conventional $\alpha$ prescription)
axisymmetric 3D disc. This study revealed the presense of a steady meridional flow
pattern with backflows for values of the $\alpha$ less than some critical value.
This result and feature was confirmed by Regev \& Gitelman (2002),
who abandoned the polytropic assumption and
included an energy equation (employing the diffusion approximation
in the treatment of the vertical radiative transport).
Umurhan \& Shaviv (2005) allowed for time dependence
and found {\em algebraic} temporal growth of global axisymmetric
adiabatic disturbances, in a polytropic inviscid disc.

The purpose of the present paper is to pursue this global asymptotic
approach to pursue the time-dependent response of a general
{\em viscous} thin accretion ($\alpha$) disc and to identify the terms
(mechanisms) responsible for the assortment of resulting 
behaviour, including any transient growth.
The main departure that this work undertakes from previous
investigations of TG in accretion discs
(except that of Ionannaou \& Kakouris, 2001 and  Umurhan \& Shaviv, 2005)
is that we treat the dynamics of a sizeable part of the disc
rather than a very small section centered about the disc's midplane
(the shearing-box, e.g., Goldreich \& Lynden-Bell, 1965;
Balbus, Hawley \& Stone, 1996),
for which the boundary conditions are, strictly speaking, not known.
In contrast to Ionannou \& Kakouris (2001) we investigate here
3D (albeit axisymmetrical) disturbances and, as said before,
this study generalizes the work of Umurhan \& Shaviv (2005)
to viscous flows and, as such, reveals the full range of possibilities.
As we said, the primary tool utilized here in order to facilitate
an analytical treatment is the asymptotic expansion, where
the dependent variables and governing equations
are expanded in powers of
some small quantity (here the measure of the disc's `thinness', $\epsilon$).
Exposing the resulting mathematical system to a set of perturbations,
in which the extreme geometry of the disc structure is taken into account
(i.e. its 'thinness'),
we aim at obtaining an analytically treatable problem.
To achieve this goal we also assume a polytropic relation between the
pressure and density.
Since this assumption was found to make little
substantive difference to the {\em steady} meridional flow solution
(Regev \& Gitelman, 2002, and cf. Klu\'zniak \& Kita, 2000),
we see its use as justified here as well.
The standard $\alpha$
prescription for the viscosity will be used here and thus we actually will
be dealing with flows whose effective Reynolds number is
${\rm Re} \sim 1/\alpha$.

The advantage of the asymptotic method and approach and the simplistic
polytropic assumption is obvious - the
treatment will be analytical and the responsible physical effects
leading to any interesting dynamics may be transparently traced.
It is clear, however, that the present analytic analysis should be
ultimately complemented with a detailed and uncompromising numerical
solution and a proper treatment of energy generation and transfer.

As we shall see, the inviscid algebraic growth found
in Umurhan \& Shaviv (2005) will be replaced in this general viscous case
by transient growth. The TG maximal magnitude will be seen to
depend on $\alpha$ (i.e. the Reynolds number) and on the nature of
the initial disturbance. It is still an open question as to 'if' and 'how' non-linearities
exploit the transiently growing disturbances to drive sustained turbulence.
Given that this work is a linear study, this matter is not the subject of our discussions.
We would like to remark, however, that
repeated (every few hundred rotations or so) perturbative events can
do the job of keeping dynamical activity alive (e.g., Ioannaou \& Kakouris, 2002)
and such events can not be ruled out in a real astrophysical
accretion disc (e.g. due the presence of the second star in binary systems,
variability of the circumstellar environment in young stars,
various dynamical events in the vicinity of an AGN etc.).

This paper is organized in the following way.
In Section 2 we state all our assumptions,
derive the basic non-dimensional equations
and introduce the asymptotic expansions for the dependent variables
that are used in this work. Although
we are bound to repeat here some previously published
work, we feel that explaining the global asymptotic approach to thin
accretion discs (here including also time dependence) in a self-contained way
may be useful for better understanding of our work here as well as
perhaps for future investigations.
In Section 3 the steady solutions of the equations are given and discussed.
These solutions are then used to study the dynamical evolution
by adding (in a suitable order) a time dependent disturbance to them.
The dynamics of the disturbances is the subject of Section 4, where we present
the results and discuss their nature. A global principle as well as
detailed dynamical behaviour and some limits are discussed in an effort
to uncover the origin of our primary result - transient growth.
The case of a general polytropic index and some technical details
of the analytical asymptotic procedure needed for obtaining the solutions
are given in the Appendices.

 \section{Assumptions, non dimensional equations and asymptotic expansions}
 Our starting point are the general Navier-Stokes equations
 \beqa
 \rho \frac{\partial{\bf V}}{\partial t} + \rho ({\bf V}\cdot\nabla) {\bf V} &=&
 -\nabla P + \rho {\bf b} + \nabla\cdot{\bsigma},
 \label{full_momentum} \\
\frac{\partial{\rho}}{\partial t} + \nabla\cdot(\rho \bf V) &=& 0,
 \eeqa
where ${\bf V}$ is the three dimensional velocity vector, $\rho$ is the density,
$P$ is the pressure and ${\bf b}$ is the body force per unit mass.
The Cartesian components of the viscous stress tensor, ${\bsigma}$ are given by
\beq
\sigma_{jk} = \eta\left( \frac{\partial V_j}{\partial x_k} + \frac{\partial V_k}{\partial x_j}
- \frac{2}{3}\delta_{jk} \nabla\cdot{\bf V} \right ) ,
\eeq
in which $\eta$ is the dynamic viscosity coefficient (we neglect bulk viscosity because
the processes discussed here are sufficiently slow to take place under thermodynamic equilibrium).

In the context of accretion discs it is natural to present the equations
in cylindrical
polar coordinates with $r$, $z$ and $\phi$ being the radial, vertical
and azimuthal coordinates
respectively. Also, the disc matter's self gravity is neglected and therefore
the body force derives from the gravitational potential of a central accreting
object, whose mass is $M$, say. Thus we have ${\bf b} = -\nabla \Phi$ with
\beq
\Phi  =  - \frac{GM}{\sqrt{r^2 + z^2}}.
\eeq
Regarding the viscosity of the accretion disc flow, we shall assume that
the situation is such that the viscosity coefficient is greatly enhanced
relatively to the microscopic one (see above in the Introduction).
We will thus use one of the standard
$\alpha$-prescriptions (see below). If this effective viscosity enhancement results
from $\eta$ actually being the eddy viscosity of an already turbulent flow,
the variables of the fluid dynamical equations should be understood
as {\em mean} quantities obtained by Reynolds averaging, a standard
technique in treatments of turbulence (e.g. Monin \& Yaglom, 1971).

We proceed in a manner similar to that of  Klu\'zniak \& Kita (2000), hereafter KK, and
Regev \& Gitelman (2002), hereafter RG, by writing the equations in cylindrical coordinates
(with the assumption of axisymmetry) in their non-dimensional form, which allows
analytical asymptotic treatment.
The difference is that here we shall not assume a steady flow:
we retain the time derivative terms. This will enable us to analyze the dynamical
evolution of deviations from the steady KK solution.

\subsection{The scalings and additional assumptions}

We consider here cold discs.  Physically this means that the characteristic disc height
(as measured from the midplane) is much smaller than its characteristic radius as measured
from the accreting star's center. This feature is readily brought to the fore when
we scale the dependent variables of the problem by their characteristic values
(see Regev, 1983, hereafter R), which we shall denote by the `tilde' sign.
With $\tilde R$ the characteristic radial scale of the disc
(which is also a natural unit of the radial coordinate),
the natural scaling for the angular velocity is
the Keplerian value $\tilde \Omega \equiv
\Omega_k(\tilde R) = (GM/\tilde R^3)^{1/2}$. Also,
the density will be scaled by a characteristic value (say that at the radius $\tilde R$
and at the midplane of the disc): $\tilde \rho$.

We assume throughout this work
that the disc equation of state is polytropic - both in steady and dynamical
states. This means that we can assume that the pressure and density
always obey the relationship
$P = P(\rho) \equiv K  \rho^{(1+1/n)}$, where $n$ is the (constant) polytropic
index and $K$ is a constant. It follows that the typical scale of the pressure is
$\tilde P = P(\tilde \rho)$.
Also, we choose the typical sound speed to be $\tilde c_s =
\sqrt{\tilde P/\tilde \rho}$.
The assumption that the disc is {\em cold} means that $\tilde c_s \ll \tilde R \tilde \Omega$
which, in turn, allows for the definition of the vertical scale height
of the disc, $\tilde H \equiv \tilde c_s \tilde \Omega$.
This is thus the natural\footnote{Note that though a `cold' disc implies
$H/R \ll 1$, a `hot' disc, i.e.
one where $\epsilon \gg 1$, does {\em not imply} that $H/R \gg 1$.  Instead it just
means that the structure starts looking more like a star with $H/R \sim 1$.} scale
for the coordinate $z$.
Formally speaking, we may define the small expansion parameter $\epsilon$ to be,
\beq
\epsilon \equiv \frac{\tilde c_s}{\tilde R  \tilde \Omega} = \frac{\tilde H}{\tilde R}
\ll 1.
\label{definition_of_epsilon}
\eeq
This disparity of scales is exploited in what follows.
The radial ($V_r$) and vertical ($V_z$) velocities are scaled by the
typical sound speed $\tilde c_s$ and
the angular velocity ($\Omega = V_{\phi}/r$) by $\tilde\Omega$.
Furthermore, we find that there is one natural choice for the temporal
scaling, since a typical rotation time $\tilde\Omega^{-1}$
and the vertical sound crossing time, $\tilde H / \tilde c_s$,
are defined to be identical.

>From here on  all variables are assumed to be
non-dimensionalized according to what has just been described.
Thus the Keplerian angular velocity
and the polytropic relations for the pressure and sound speeds are
\beq
\Omega_k = \frac{1}{r^{3/2}}, \qquad
P = \rho^{1 + 1/n}, \qquad
c_s^2 = \frac{d P}{d \rho} = \left(1+\frac{1}{n}\right) \frac {P}{\rho}=
\left(1+\frac{1}{n}\right) \rho^{1/n}.
\label{polytropicrel}
\eeq
To be consistent with previous works (e.g. KK)
we make use of the non-dimensional sound speed $c_s$
as the dependent variable instead of the pressure $P$.   As such it readily follows that
\beq
\frac{1}{\rho}\frac{\partial P}{\partial r} = n\frac{\partial c_s^2}{\partial r},\qquad
\frac{1}{\rho}\frac{\partial P}{\partial z} = n\frac{\partial c_s^2}{\partial z}.
\eeq

The standard $\alpha$ model of Shakura \& Sunyaev (1973)
is based on the assumption that the only non vanishing viscous stress component
is $\sigma_ {r \phi}$ and it is proportional to the pressure.
Following KK, we adopt this assumption and derive from it the form of
viscosity coefficient, but include in the dynamical equations all the
components of the stress tensor.
In lowest order in $\epsilon$
the angular velocity of a disc is Keplerian and  we get
(with the dynamic viscosity coefficient scaled by $\tilde c_s \tilde H$)
the nondimensional relation (see also RG)
 \beq
 \eta  = \frac{2}{3}\frac{\alpha c_s^2}{\Omega_k \left(1+\frac{1}{n}\right)},
 \label{full_definition_eta}
 \eeq
 where $\alpha$ is the viscosity parameter.

 \subsection{Nondimensional equations and asymptotic expansions}
 As stated in the Introduction, we seek  axisymmetric solutions
 (in both the steady and dynamical cases) of the equations of motion.
Denoting the non-dimensionalized radial and vertical velocities as $u$ and $v$,
respectively,
the equations now appear as they do in KK and RG except that here
we allow for time-dependence (the time unit is $1/\tilde \Omega$).
Consequently time-derivatives are included and all the dependant variables
 are functions of $r,z$ and $t$ -
 \beqa
\epsilon\frac{\partial u}{\partial t} + \epsilon^2 u \frac{\partial u}{\partial r}
+ \epsilon v\frac{\partial u}{\partial z} - \Omega^2r &=&
- \frac{1}{r^2}\left[1 + \epsilon^2\frac{z^2}{r^2}\right]^{-3/2}
+ \frac{\epsilon}{\rho}\frac{\partial}{\partial z}\left(\eta\frac{\partial u}{\partial z}\right) \nonumber \\
&&
+ \epsilon^2 \left[-n \frac{\partial c_s^2}{\partial r} + \frac{1}{\rho}\frac{\partial}{\partial z}
\left(\eta \frac{\partial v}{\partial r}\right) -
\frac{1}{\rho}\frac{\partial}{\partial r}
\left(\eta\frac{2}{3} \frac{\partial v}{\partial z}\right)
\right]
\nonumber
\\ & & + \epsilon^3\left[-\frac{2\eta u}{\rho r^2} + \frac{1}{\rho r}\frac{\partial}{\partial r}
\left(2\eta r \frac{\partial u}{\partial r}\right)  - \frac{1}{\rho}\frac{\partial}{\partial r}
\left(\frac{2}{3}\frac{\eta}{r}\frac{\partial(ru)}{\partial r}\right)   \right],
\label{expanded_u_eqn}\\
\rho\frac{\partial\Omega}{\partial t} + \epsilon\frac{\rho u}{r^2}\frac{\partial (r^2 \Omega)}{\partial r}
+\rho v\frac{\partial\Omega}{\partial z} &=&
\frac{\partial}{\partial z}\left(\eta\frac{\partial \Omega}{\partial z}\right)
+\epsilon^2 \frac{1}{r^3}\frac{\partial}{\partial r}
\left(\eta r^3 \frac{\partial \Omega}{\partial r}\right)
\label{expanded_Omega_eqn}\\
\frac{\partial v}{\partial t} + \epsilon u \frac{\partial v}{\partial r}
+v\frac{\partial v}{\partial z} &=&
- \frac{z}{r^3}\left(1 + \epsilon^2\frac{z^2}{r^2}\right)^{-3/2}
-n \frac{\partial c_s^2}{\partial z} +
\frac{4}{3}\frac{1}{\rho}\frac{\partial}{\partial z}\left(\eta\frac{\partial v}{\partial z}\right)\nonumber \\
& & + \epsilon\frac{1}{\rho}\left[
\frac{1}{r}\frac{\partial}{\partial r}\left(\eta r\frac{\partial u}{\partial z}\right)
- \frac{2}{3}\frac{\partial}{\partial z}\left(\frac{\eta}{r}\frac{\partial (ru)}{\partial r}\right)
\right]
+\epsilon^2\frac{1}{\rho r}
\frac{\partial}{\partial r}\left(\eta r\frac{\partial v}{\partial r}\right),
\label{expanded_v_eqn}\\
\frac{\partial\rho}{\partial t} + \frac{\epsilon}{r}\frac{\partial (r\rho u)}{\partial r}
+ \frac{\partial (\rho v)}{\partial z} & = & 0,
\label{expanded_continuity_eqn}
 \eeqa
 where the gravitational potential of the central star has been expanded only up to
 second order in $\epsilon$.

 We seek asymptotic solutions to (\ref{expanded_u_eqn}-\ref{expanded_continuity_eqn})
 by writing all the dependent variables in a power series of
the small parameter $\epsilon$, i.e. we formally write
 \beqa
 {\bf U}(r,z,t) = \sum_{m=0}^{\infty} \epsilon^m\, {\bf U}_m,
 \eeqa
where ${\bf U} = (u,v,\Omega,c_s^2)^{{\bf T}}$.  We remind
the reader that $c_s$ and $\rho$ are not
independent (on account of the assumption of polytropes)
and are related through (\ref{polytropicrel}).

It has been shown before (see R, KK, RG) that in steady state
the lowest order nonzero components of the meridional velocities
are $u_1$ and $v_2$ and, in addition, the choice
$\Omega_1= \rho_1=0$ (and thus also $c_{s1}=0$) can be consistently made.
Guided by these results (see also below) we retain in the
expansions for $\Omega$, $c_s^2$, $\rho$ and $v$ only even powers
of $\epsilon$ (in the case of $v$ starting from $v_2$), while
for the $u$ expansions only odd powers are kept, starting with $u_1$,
 \beqa
 \Omega(r,z,t) &=& \Omega_0(r,z) +\epsilon^2\left[\Omega_2(r,z) + \Omega_2'(r,z,t)\right] +
 \epsilon^4\left[  \Omega_4 +\Omega_4'(r,z,t)\right]+\cdots
 \label{expan1}\\
 u(r,z,t) &=& \epsilon \left[u_1(r,z) + u_1'(r,z,t)\right] +  \epsilon^3 \left[ u_3(r,z) +u_3'(r,z,t)\right]+ \cdots \\
 v(r,z,t) &=& \epsilon^2\left[v_2(r,z) + v_2'(r,z,t)\right] + \epsilon^4 \left[v_4(r,z) + v_4'(r,z,t) \right] + \cdots \\
 c_s^2(r,z,t) &=& c_{s0}^2(r,z) + \epsilon^2\left[c_{s2}^2(r,z) + {c_{s2}^2}'(r,z,t)\right]
 + \epsilon^4\left[ c_{s4}^2(r,z) + c_{s4}' (r,z,t)\right] + \cdots \\
 \rho(r,z,t) &=& \rho_0(r,z) + \epsilon^2\left[\rho_2(r,z) + \rho_2'(r,z,t)\right]
 + \epsilon^4 \left[ \rho_{4}(r,z) + \rho_4'(r,z,t)\right]+ \cdots
 \label{expan5}
 \eeqa
All  $\order{1}$ quantities are assumed to be time-independent
(in accord with the steady state results of RG and KK).
As is also apparent, we have assumed
no meridional flow at $\order{1}$ since,
as said above, it has been shown before that $u_0=v_0=v_1=0$.
Time dependence has been introduced into the expansions
at $\order{\epsilon^2}$ for the functions
 $\Omega,\rho, c_s^2$ and $v$ while it appears at $\order{\epsilon}$ for $u$.
 At all orders in which time dependence is introduced we
 split the solutions up into a sum of a steady solution and a dynamical one,
 denoted by a prime, and this is true for all high orders as well.
Thus, from now and on all the terms of the dependent variables expansions
that are dependent on time are denoted by a prime, while those without
prime are just space dependent, that is, steady.
The implicit assumption in this kind of splitting is that the time-dependent
part is a perturbation (not necessarily infinitesimal!)
on the steady state. In any case, the expansions Ansatz (\ref{expan1})-(\ref{expan5})
can always be considered as a particular {\em choice} of a perturbation
kind.

The steady part of the solutions up to second order
will be identical to the solutions obtained in KK and RG (for $n=3/2$).
In this work we shall limit the analysis of the evolution of the time-dependent
disturbances to terms of order $\epsilon^2$ or lower.
 \section{Steady State}
 \subsection{$\order 1$}
All that is different between the results given this section and
the results of KK is that all expressions are derived here in terms of arbitrary polytropic
index $n$ while in KK the specific value of $n=3/2$ is assumed throughout their study.
The lowest order equations are straightforwardly solved to yield
solutions quite well-known in the literature (see, e.g., Hoshi, 1977).
These are,
\beq
 c_{s0}^2(r,z)  = \frac{h^2(r) - z^2}{2nr^3}, \qquad
 \Omega_0 = \Omega_k =\frac{1}{r^{3/2}},  \qquad
 \rho_0(r,z)  = \left[\frac{c_{s0}^2(r,z)}{1+\frac{1}{n}}\right]^n.
 \label{order_zero_steady_state}
 \eeq
 The height of the disc $h(r)$, i.e. where the quantities $c_{s0}^2$ and $\rho_0$ go to zero,
 is explicitly determined when solving the next order steady state equations.  For values of
 $|z| \ge h$, $c_{s0}^2$, $\rho_0$ and all other associated quantities remain zero.
 Additionally the relationships between the sound speed, pressure and density, as well
 as their vertical gradients are given by,
 \beq
 \frac{P_0}{\rho_0} = \frac{c_{s0}^2}{1+\frac{1}{n}},\qquad
 \frac{1}{\rho_0}\frac{\partial P_0}{\partial z} = n\frac{\partial {c_{s0}^2}}{\partial z} =
 -\frac{z}{r^3}.
 \label{polyrelations}
 \eeq
 It useful to note that we shall use in the equations to all orders
 only the zeroth order value of the viscosity
 coefficient, that is, will not actually consider it as
 an exapandable function, but rather as a prescribed function
 which is based on the zeroth order of the pressure.
Thus
 \beq
 \eta \equiv \eta_0= \sfrac{2}{3}\alpha P_0 r^{3/2}.
\label{eta0_P0_relationship}
 \eeq
The steady state at this order of $\epsilon$ has obviously a vanishing meridional flow
(see KK).

 \subsection{$\order{\epsilon^2}$}
 \label{ss}
 The steady-state equations at the next non-trivial order of $\epsilon$   are,
 \beqa
 n\frac{\partial c_{s0}^2}{\partial r} - \frac{3}{2}\frac{z^2}{r^4}
  &=& 2\Omega_0\Omega_2r + \frac{1}{\rho_0}\frac{\partial}{\partial z}\left(
 \eta \frac{\partial u_1}{\partial z}\right),
 \label{SS_epsilon2_ueqn}
\\
-\frac{1}{r^3 \rho_0}\frac{\partial}{\partial r}\left(
 \eta r^3\frac{\partial \Omega_0}{\partial r}\right)
&=& -\frac{ u_1}{r^2} \frac{\partial (r^2\Omega_0)}{\partial r} + \frac{1}{\rho_0}\frac{\partial}{\partial z}\left(
 \eta \frac{\partial \Omega_2}{\partial z}\right),
 \label{SS_epsilon2_Omegaeqn}
\\
 0 &=& \frac{1}{r}\frac{\partial}{\partial r} (r\rho_0 u_1) + \frac{\partial}{\partial z} (\rho_0 v_2),
\label{SS_epsilon2_continuityeqn}
\\
0 & = &  \frac{3}{2}\frac{z^3}{r^5}
-n \frac{\partial c_{s2}^2}{\partial z} +
\frac{4}{3}\frac{1}{\rho_0}\frac{\partial}{\partial z}
\left(\eta \frac{\partial v_2}{\partial z}\right)
+\frac{1}{\rho_0}\left[
\frac{1}{r}\frac{\partial}{\partial r}\left(\eta r\frac{\partial u_1}{\partial z}\right)
- \frac{2}{3}\frac{\partial}{\partial z}\left(\frac{\eta }{r}\frac{\partial (ru_1)}{\partial r}\right)
\right].
\label{SS_epsilon2_veqn}
 \eeqa
Inspection of
these equations show that they contain as solutions meridional flow,
something which we will formalize below.
\par
KK found analytically the solutions
at this order (i.e. equations  \ref{SS_epsilon2_ueqn} - \ref{SS_epsilon2_veqn})
for the polytropic index $n=3/2$. We discuss in considerable detail the
solutions for an arbitrary polytropic index $n$ in Appendix
\ref{steady_state_KKsolutions}.
Here we would like only to stress some important points.

First, as KK demonstrated,
there is a value of $\alpha$ below which there is a
certain amount of backflow in the disc, that is,
a stagnation radius exists, beyond which
the radial velocity, near the disc midplane, is directed outwards.

Second, also as KK demonstrated,
the disc height as a function of $r$, $h(r)$,
depends on the mass accretion rate, which in the appropriate
units turns out to be of order $\epsilon$, $\dot M = \epsilon \dot M_1$,
and the polytropic index $n$.
The disc's height $h(r)$ can be derived in an asymptotically valid way
only for radii significantly larger than $r_*$, where
$r_*$ is the {\em zero torque} position, i.e. the point at which
the angular velocity (having the Keplerian value at lowest order) must
pass a maximum on the way to matching it to some inner boundary value
(the angular velocity of the surface of the accreting object, which
is sub-Keplerian). The point, if treated carelessly, will produce pathologies
like zero disc thickness, a zero density and divergent velocities.
To find the structure near $r_*$, some boundary layer treatment,
using singular perturbation methods like the ones employed by, for example, Bertout \& Regev (1995)
(see also references therein) or fully numerical approaches (see Popham \& Narayan, 1995;
and references therein), must be employed. Thus in this work, as well as
in KK and RG, we are dealing with the disc only for radii sufficiently
larger than $r_*$.

Third, these solutions do not allow for a detailed outer boundary condition
(only an integral condition for the constancy of the total mass influx in the disc
is employed).
Thus, the KK solution and our generalization thereof (both steady and dynamical)
are valid in regions of the disc that are not too close to its edges in the
radial direction.

Finally, it is an interesting result that in the expression for the disc
height for arbitrary polytropic index (see \ref{disc_height})  the disc
flares for  $n > 3/2$, namely $h(r) \sim r^{m}$ with $m > 1$.

The form of $h(r)$ given in KK is recovered by substituting
$n=3/2$ into the general expressions (see \ref{disc_height}) given in Appendix A and reads
as follows,
\beq
h(r) =
(2\Lambda)^{\frac{1}{6}}r^{\frac{11}{12}}
\left(\sqrt r - \sqrt r_*\right)^{\frac{1}{6}},\qquad
\Lambda \equiv \frac{\dot M_1}{\alpha}\frac{\Gamma(4)}{\Gamma(5/2)}
\cdot \frac{(5)^{\sfrac{3}{2}}}{3\pi^{3/2}(11/2)}.
\label{h_n3halves}
\eeq
The height of the disc at the fiducial radius $r=1$ is denoted by $h_1$ and is
\beq
h_1 \equiv
(2\Lambda)^{\frac{1}{6}}
\left(1 - \sqrt r_*\right)^{\frac{1}{6}}.
\label{h1_n3halves}
\eeq
\par

 \section{Dynamical evolution}
 The expansion procedure outlined in the previous section for the steady state solutions is
 extended here to  dynamical problems.  In other words, we dynamically evolve
 all equations at successive
 orders of $\epsilon$ subject to the appropriate boundary conditions.
 Regarding the radial boundary conditions, we avoid the problems posed by them by
 considering only the portion of the disc with
 an inner radius $r_{\rm in}$ and an outer
radius $r_{\rm max}$, as in the steady-state case (see KK).

In the following, we discuss the vertical boundary conditions which we adopt.
Let us say from the outset that the choices we make
seem to us to be physically reasonable and, yet, allow for some analytical tractability.
In steady-state, the  disc lies between $z = \pm h(r)$, in which $h(r)$ is as
given in the previous sections.  In this work we this consider the `surface' of the
disc to consist of the last fluid parcel which, in a steady disc, is at $z = h(r)$.
Technically speaking, the surface of the dynamically evolving
disc should be given by $h(r) + \order{\epsilon^2}$, where the correction term accounts
for the evolving surface.
However, to the order up to which equations are sought in this work, it suffices to enforce
the boundary conditions at $z=h(r)$.\footnote{In other words, the 
$\order{\epsilon^2}$ correction term to the disc surface
does not alter the resulting disc dynamics at the
lowest order for which they are determined.}
In addition, the equatorial dynamical behaviour allows two kinds of dynamical considerations
according to whether the disturbances
in the disc are {\em sinuous} or {\em varicose}. Varicose modes have even symmetry about the $z=0$
axis in all the quantities except the vertical velocity which has odd symmetry there.
Sinuous modes have opposite symmetry of varicose modes.

Firstly, thus, in this work we consider only varicose modes (the steady state meridional
flow has this symmetry as well).
However, it should be noted that when mode couplings begin to play a role
in the evolution of the system (which occur at orders of $\epsilon$ higher than
that explored in this work), sinuous and varicose modes dynamically influence
each other via the non-linear advection terms in the equation for
the vertical velocity, that is,
the $v\partial_z v$ term in (\ref{expanded_v_eqn}) provides this cross-mode
interaction.

Secondly, we allow the disc surface to flutter about freely in a Lagrangian sense.  This, in turn,
means that we require that the pressure on the moving
surface vanish for all times.  This, then, leads
to the condition that the
Lagrangian pressure fluctuation vanishes on this surface i.e.,
\beq
\left(\frac{\partial}{\partial t} + \epsilon u\frac{\partial}{\partial r} + v\frac{\partial}{\partial z}
\right)P = 0, \ \ \ {\rm at} \ \ z = \pm h(r) .
\nonumber
\eeq
Given the polytropic relation (\ref{polytropicrel}), the above boundary condition
may be simplified with the aid of the mass continuity equation
(\ref{expanded_continuity_eqn}), to 
\beq
P\left(\epsilon\frac{\partial u}{\partial r} + \frac{\partial v}{\partial z}\right)
= 0 \ \ \  {\rm at} \ \  z = \pm h(r).
\label{lagrangian_pressure_BC}
\eeq
The above is  a restatement of the condition
that there should be no thermodynamic work performed at the moving boundary.
\par
Thirdly, we posit that the viscous stresses on the surface of the disc
are zero.  This means that the product of the viscosity coefficient
and the velocities (or the velocity gradients) tend to zero on a boundary.
Since this is a free surface and the viscosity prescription given tends
to zero at zero density, we will impose the more restrictive free-surface condition,
\beq
\lim_{z\rightarrow h} \
\eta r {{\bf \hat n}}\cdot{\nabla}\Omega   \rightarrow 0 \ \  \ {\rm and} \ \ \
\eta {{\bf \hat n}}\cdot{\nabla u}  \rightarrow 0,
\eeq
where ${{\bf \hat n}}$ is the normal to the surface ${\bf S}$ of the disc.
In the event where the
surface of the disc is approximately aligned with the normal to the
vertical direction this condition becomes,
\beq
\lim_{z\rightarrow h} \
\eta r\frac{\partial \Omega}{\partial z}  \rightarrow 0  \ \ {\rm and} \ \
\lim_{z\rightarrow h}
\eta\frac{\partial u}{\partial z}  \rightarrow 0.
\eeq
In other words, we allow the gradient of the $u$ and $r \Omega$ velocities to diverge
as one approaches the disc height $h$ but the rate thereof should be slower than
the rate at which $\eta$
tends to zero. In the solutions
that are developed for these modes (see below),
we find that this condition is always satisfied
and there are never diverging gradients of either $u$ or $\Omega$.
\par
Lastly,
we impose that the total vertically integrated mass flux through an annular section of the
disc be fixed and steady.  In turn this means that the vertically integrated mass flux through an
annular disc section {\em of dynamically
varying quantities} ought to be identically zero, i.e.
\beq
2\pi r\int_{-\infty}^{\infty}{\rho  u'_i}dz = 0,
\eeq
in which $u'_i(r,z,t)$ are the dynamically varying portions of the radial velocity
at each order $i$ of the $\epsilon$ expansion.

  \subsection{$\order{\epsilon^2}$ dynamical equations}
  With the solution expansions implemented into the equation set
  (\ref{expanded_u_eqn}-\ref{expanded_continuity_eqn}) we find
  at the lowest nontrivial order describing the dynamical evolution
  of the disturbances (denoted by the primed quantities)
   the following,
 \beqa
 \partial_t u_{1}' &=& 2\Omega_0\Omega'_{2}r +
\frac{1}{\rho_0}\frac{\partial}{\partial z}\left(
 \eta \frac{\partial u'_{1}}{\partial z}\right),
\label{u_mom_eq}
\\
 \partial_t \Omega'_{2} &=& -\frac{ u'_{1}}{r^2} \frac{d (r^2\Omega_0)}{d r}
 + \frac{1}{\rho_0}\frac{\partial}{\partial z}\left(
 \eta \frac{\partial \Omega'_{2}}{\partial z}\right),
\label{Omega_mom_eq}
\\
 \partial_t v'_{2} &=&
 -n\frac{\partial {c_{s}^2}'_{2}}{\partial z}
 +\frac{4}{3}
\frac{1}{\rho_0}
 \frac{\partial}{\partial z}\left(\eta
 \frac{\partial v'_{2}}{\partial z}\right) +
\frac{1}{r}\frac{1}{\rho_0}\frac{\partial}{\partial r}
\left(\eta r\frac{\partial u'_{1}}{\partial z}\right)
 -  \frac{2}{3}
\frac{1}{\rho_0}
 \frac{\partial}{\partial z}\left[\eta
 \left(\frac{1}{r}\frac{\partial (ru'_{1})}{\partial r}\right)\right],
\label{v_mom_eq} \\
 \partial_t \rho'_{2}
&=& -\frac{1}{r}\frac{\partial}{\partial r} (r\rho_0 u'_{1})
- \frac{\partial}{\partial z} (\rho_0 v'_{2}).
\label{continuity_eq}
 \eeqa
 The polytropic equation state tells us that the relationship between the second
 order density and sound speed dynamical disturbances is
 \beq
 {c_{s}^2}'_{2} = c_{s0}^2 \frac{1}{n}\frac{\rho'_{2}}{\rho_0}.
 \label{relation_of_rhoprime_csprime}
 \eeq
 This algebraic relation is necessary to complete the linear PDE set
(\ref{u_mom_eq})-(\ref{continuity_eq}).

It is important to note that the above four equations support
two sorts of modes, namely a pair
comprising of purely vertical disturbances and a
second pair containing disturbances in all the velocity
components.
The nature of these modes is detailed below. Such
a classification is possible because
equations (\ref{u_mom_eq}-\ref{Omega_mom_eq})
dynamically decouple from
(\ref{v_mom_eq}-\ref{continuity_eq}),  i.e. the quantities
$u'_{1}$ and $\Omega'_{2}$ are not dynamically influenced by $v'_{2}$
and $\rho'_{2}$, while the converse is not true.
That is, disturbances in $u'_{1}$ and $\Omega'_{2}$ will
drive a dynamical response in $v'_{2}$ and $\rho'_{2}$. This mathematical
structure is reminscent of the behaviour of the steady-state equations (see
section \ref{ss} and KK).

To be more specific (\ref{u_mom_eq}-\ref{Omega_mom_eq})
and (\ref{v_mom_eq}-\ref{continuity_eq}) may be combined
to give two second order equations for $u'_{1}$ and $v'_{2}$,
\beqa
{\cal P} u'_{1} &=& 0,
\label{single_u_eqn}\\
{\cal L}v'_{2}&=& \left[\partial_t {\cal F} +  {\cal G} \right]u'_{1},
\label{combo_v_eq}
\eeqa
in which the operators are,
\beqa
{\cal P} &\equiv&
\left(\partial_t  - \frac{1}{\rho_0}\frac{\partial}{\partial z}\eta \frac{\partial}{\partial z}
\right)^2 + \Omega_0^2,
\label{P_def}\\
{\cal L} &\equiv& \partial_t^2  - c_{s0}^2\frac{\partial^2}{\partial z^2}
-(n+1)\frac{\partial c_{s0}^2}{\partial z}\frac{\partial }{\partial z}
-n \frac{\partial^2c_{s0}^2}{\partial z^2}
-\frac{4}{3} \frac{1}{\rho_0}\frac{\partial}{\partial z}
\eta \frac{\partial}{\partial z} \partial_t,
\label{L_def} \\
{\cal F} &\equiv&
-\frac{2}{3 }\frac{1}{r \rho_0}\frac{\partial}{\partial z}\eta \frac{\partial }{\partial r}r
+
\frac{1}{r\rho_0}\frac{\partial}{\partial r}{r\eta}\frac{\partial }{\partial z},
\\
{\cal G} &\equiv& \frac{\partial}{\partial z}
\frac{c_{s0}^2}{r\rho_0}\frac{\partial  }{\partial r}\rho_0 r.
\eeqa

The zero Lagrangian pressure condition gives, at this order,
\beq
P_0\left(\frac{\partial u'_{1}}{\partial r}
+ \frac{\partial v'_{2}}{\partial z} \right)= 0, \ \ {\rm at} \ \  z=h(r),
\label{order_epsilon2_pressurecondition}
\eeq
 which, we must reiterate, is at the unperturbed disc boundary.
Also, the stress conditions are
\beq
\eta r\frac{\partial \Omega_2'}{\partial z}  = 0  \ \ {\rm and} \ \
\eta\frac{\partial u_1'}{\partial z}  =0
, \  {\rm at}\  z=h(r).
\label{order_epsilon2_stresscondition}
\eeq
Finally,
the integrated mass flux condition is equivalent to requiring
\beq
\label{eq:zeromass}
2\pi r \int_{-h}^{h} {\rho_0 u'_{1}}dz = 0.
\eeq

We turn now to the presentation of analytic solutions of the system
(\ref{single_u_eqn}-\ref{combo_v_eq}).
We shall first discuss, in section \ref{homo}, the case $u_1'=0$
(and thus also $\Omega_2'$=0),
i.e. solving only the homogeneous part of
(\ref{combo_v_eq}) . The solutions of
this equation constitute the above mentioned first pair of
modes (purely vertical disturbances) and we shall refer to them
as {\em vertical acoustics}.
The solutions to the full set (\ref{single_u_eqn}-\ref{combo_v_eq}) in
the general case will be discussed in section \ref{general}. This
pair of modes will be referred to as {\em driven general acoustics}.
The transient growth behaviour found in our solutions will be detailed
in section \ref{TG}.
A full exposition of the calculations leading to the solutions quoted
and discussed in sections
\ref{homo}-\ref{TG} is presented in
in Appendices \ref{homo_acoustics_details}-\ref{inertio_viscous_modes}.
In the next three subsections only the salient features will be given.

\subsection{Results: (i) Vertical acoustics}
\label{homo}
The general solution to the inhomogeneous equation (\ref{combo_v_eq}) can be written,
in the form
\begin{equation}
v'_{2}=v'_h + v'_p
\label{vseparation}
\end{equation}
where the index $h$ stands for a solution of the homogenous equation and the index
$p$ stands for particular solution of the inhomogenous equation.
As we have remarked earlier,
the interesting property of the equations, which is a result of the disc being
geometrically thin, is that the velocity and density fluctuations do not induce
neither radial nor azimuthal motions at these lowest orders.
This implies that homogeneous solutions of (\ref{combo_v_eq}),
(i.e. with the equation's RHS set to zero, because $u_1'=\Omega_2'=0$ say,
see above) are perfectly acceptable.

These solutions have the form (see Appendix \ref{homo_acoustics_details})
\beq
v_h' = \hat v_h'(z,r)\exp(\tilde\sigma T) + {\rm c.c.} \ \ \ {\rm with} \ \ \
T\equiv{t \over r^{3/2}},
\label{homo_acoustics_ansatz}
\eeq
where the spatial eigenfunctions $\hat v_h'$ are composed of the associated Legendre functions.
This form of $v_h'$ already indicates that these solutions
are inseparable in $r$ and $t$.
The temporal
eigenvalues, which
appear in complex conjugate pairs, are functions
of $\alpha$, the polytropic index $n$ and the acoustic overtone
(labelled by the integer index $k$, say). Thus the eigenvalues of the vertical
acoustic modes should actually be written as $\tilde \sigma_k(\alpha,n)$.
For instance,
the eigenfrequency of the fundamental (i.e., $k=0$) mode, is given by
\beq
\tilde\sigma_{k=0}(\alpha,n) \equiv \tilde \sigma_F =
-\frac{4}{9}\alpha \pm  i \left| \left(\frac{16}{81}\alpha^2 -
\frac{2n+1}{n} \right) \right| ^{1/2}, \ \ \ \ 0< \alpha < 1 \ \ ; \ \ n >1.
\eeq
This shows that the fundamental mode comprises of decaying oscillations
and all overtones show such a decay as well
(see Appendix \ref{homo_acoustics_details}).
In Figure \ref{homo_acoustics_plots} we display the spatial structure of the
eigenfunctions
$\hat v_h'$ and their vertical gradient $\partial \hat v_h'/\partial z$
for the first three mode indices  and for three polytropic indices.
The structure of the eigenfunction  is found not to be very sensitive
to the polytropic index.
\begin{figure}
\begin{center}
\leavevmode \epsfysize=9.cm
\epsfbox{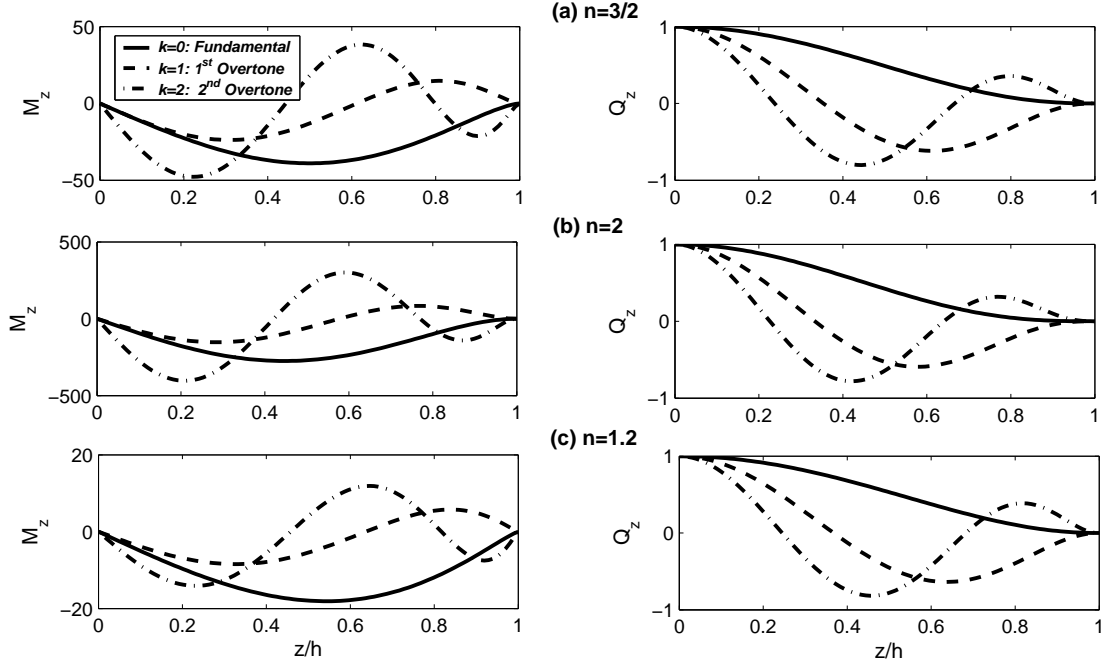}
\end{center}
\caption{{\small
Structure functions for the homogeneous acoustic modes for a few values of the polytropic
index and overtone number for values of $\alpha = 0.1$.  Plotted are the vertical mass flux
$M_z = \rho_0 \hat v_h'$ and $Q_z = P_0{\partial \hat v_h'/\partial z}$.  All $Q_z$ are scaled
to their values at $z=0$.
}}
\label{homo_acoustics_plots}
\end{figure}

\subsection{Results: (ii) Driven general acoustics}
\label{general}
We turn now to the general solution to (\ref{single_u_eqn}-\ref{combo_v_eq}) by first
focusing on (\ref{single_u_eqn}).  This equation admits an infinite set of eigenmode
solutions in a way similar in character to the homogeneous solutions
discussed in the last section.  The general derivation and structure of these eigenmodes is detailed in
Appendix \ref{inertio_viscous_modes}.  For the sake of clarity, the discussion in
the main text here shall focus mainly upon the dynamics associated with the fundamental
mode of the operator ${\cal P}$ and we shall restrict our attention to the case $n=3/2$.

The solution to the dynamical radial velocity profile in an $n=3/2$ polytrope
is given by,
\beq
u_1' = \hat u'(z,r)e^{\sigma T + \varphi} + {\rm c.c.}, \ \ \ {\rm with } \ \ \ T={t \over r^{3/2}}
\eeq
in which the spatial eigenfunction $\hat u'(z,r)$ has the particularly simple structure
\beq
\hat u' = A(r)\left(\frac{z^2}{h^2} - \frac{1}{6} \right).
\label{u_prime_solution}
\eeq
$A(r)$ is an amplitude whose radial functional form is technically arbitrary and
would be set by the initial disturbance condition. An arbitrary phase factor of $\varphi$ is
also introduced.
The eigenvalues, denoted in this case by $\sigma$, come in complex conjugate pairs given by
\beq
\sigma = \pm i - \sfrac{8}{5}\alpha.
\label{eigenvalue_inertioviscous_k0}
\eeq
Without loss of generality we need only consider  the `$+$' solution since the phase
factor $\varphi$ takes care of the `$-$' solution.
The temporal behaviour of these modes is again one of
decaying oscillations, but with frequencies
given by the local rotation rate of the disc.
Note that according to the prescription for $h$
given in (\ref{h_n3halves}),
for values of $r$ larger enough than the zero-torque radius $r_*$, the functional
dependence of the disc height with respect to $r$
is well approximated by,
\[
h(r) \sim h_1 r,\qquad \frac{dh}{dr} \sim h_1,
\]
where $h_1$ is determined by the mass flux rate and the value of $\alpha$
according to
(\ref{h1_n3halves})
or, for more general values of $n$, by
(\ref{eq:for:hr}) and (\ref{height_at_1}).  Since $h_1$ simply
adds a multiplicative factor to all the dynamical quantities, it plays no
role in the quality of the ensuing evolution and, as such, we set $h_1 = 1$
without any loss of general flavour.
\par

With this solution for $u_1'$ we may find a particular solution to the vertical
velocity, $v_p'$ by solving (\ref{combo_v_eq}) directly.  The details of this procedure
are presented in Appendix \ref{inertio_viscous_modes} but we highlight the major steps
here.\par
We posit the following form
for $v_p'$ and $\rho_p'$,
\beqa
v_p' &=& \hat v_p (\zeta,r)e^{\sigma T}
+ \hat V_{p}
(\zeta,r)\left(T e^{\sigma T}\right) \ \ +  {\rm c.c.},
\label{vp_solution_form} \\
\rho_p' &=& \hat \rho_{p}(\zeta,r)e^{\sigma T}
+ \hat R_{p}(\zeta,r)
\left(T e^{\sigma T}\right) \ \ +  {\rm c.c.},
\label{rhop_solution_form}
\eeqa
where $\zeta \equiv z/h(r)$.
Recall that we solve only for the fundamental mode of the system (again
see Appendix \ref{inertio_viscous_modes}) and that there is obviously
a dependence on $\alpha$. It is important to
notice explicit appearance of a multiplicative $T$ term in the above expression.
This temporal functional dependence is necessary in order to balance terms
arising on the RHS of (\ref{combo_v_eq}).
In general, the form of these lowest order structure eigenfunctions is
\beq
\left(
\begin{array}{c}
\hat v_{p} \\
\hat V_{p}
\end{array}
\right)
=
\left(
\begin{array}{c}
a_1 \\
b_1
\end{array}
\right)\zeta
+
\left(
\begin{array}{c}
a_3 \\
b_3
\end{array}
\right)\zeta^3,
\eeq
namely, it is polynomial in $\zeta$ of the third degree with only odd powers in $\zeta$.
The four coefficients are functions of $r$ and $\alpha$, i.e.
$a_i=a_i(r,\alpha)$ and $b_i=b_i(r, \alpha)$ for $i=1,3$.

The solution to the density perturbation, $\rho_p'$ is straightforwardly obtained
using (\ref{u_prime_solution}), (\ref{vp_solution_form}) along with $\rho_0$ defined by
(\ref{order_zero_steady_state}) and then integrating
(\ref{continuity_eq}) with respect to time. This gives

\beq
\left(
\begin{array}{c}
\hat \rho_{p} \\
\hat R_{p}
\end{array}
\right)
=
\bigl(1-\zeta^2\bigr)^{\sfrac{1}{2}}
\left[
\left(
\begin{array}{c}
c_0 \\
d_0
\end{array}
\right)
+
\left(
\begin{array}{c}
c_2 \\
d_2
\end{array}
\right)\zeta^2
+
\left(
\begin{array}{c}
c_4 \\
d_4
\end{array}
\right)\zeta^4
\right],
\eeq
where, again, the coefficients are functions of $r$ and $\alpha$, that is,
$c_i=c_i(r,\alpha)$ and $d_i=d_i(r, \alpha)$ for $i=0,2,4$.

\par
The polynomials appearing in the square
brackets of the expressions for $ \hat \rho_{p}$ and $\hat R_{p}$
have only even powers of $\zeta$. The detailed forms of the coefficients
depend on the form of the generalized initial perturbation
radial structure $A(r)$ as well as on $n$.
We avoid presenting
them here because they are very long and cumbersome; and the details of the coefficients
have no effect on the
resulting TG since they describe only a particular type of disturbance. For the sake of simplicity
all of the analytic results which will be presented below assume $A = e^{i\pi/4}$
(that is, one particular form of the initial conditions of the radial
disturbance) and $n=3/2$.

\subsection{Results: (iii) Transient growth}
\label{TG}
As is evident by inspection of (\ref{vp_solution_form}), $v'_p$ and
$\rho_p'$ exhibit TG.  To be able to present it more clearly
we consider the integral quantities

\beq
{\cal E}_a(r,t;\alpha)
\equiv h(r)\int_{-1}^{1}{\left(\sfrac{1}{2}\rho_0 v_p'^2 + \sfrac{1}{2}
\frac{c_{s0}^2\rho_p'^2}{\rho_0} \right)}d\zeta, \qquad
E_a(t;\alpha) \equiv \int_{r_{\rm min}}^{r_{\rm max}}{{\cal E}_a \ 2\pi r \ dr}.
\label{energy_a_definitions}
\eeq
The dynamical quantity ${\cal E}_a(r,t;\alpha)$ is to be interpreted as
the acoustic energy (per unit area of the disc) consisting of the kinetic
energy in the vertical velocity disturbances and the compression energy
due to the density disturbances (see Section \ref{discussion}).
The contribution of the (purely oscillatory) velocity components,
resulting from the homogeneous part the disturbance equations,
are left out when forming this energy integral.
As ${\cal E}_a$ is a function of radial position $r$, one may form
the quantity $E_a(t;\alpha)$, i.e., the {\em total disturbance
acoustic energy} (in this sense) of the disc, by integrating over
the radial range in which these disturbances were assumed
to exist, that is, between the inner and outer bounds
$r_{\rm min}$ (significantly above the inner zero-torque radius)
and $r_{\rm max}$ (significantly below the outer disc edge).

In general, both ${\cal E}_a$  and $E_a$  are  functions of the $\alpha$ parameter
as well as the structure of the radial velocity perturbation $A(r)$
(see equation \ref{u_prime_solution}).
As we have assumed (see Section \ref{general}) $A(r)$ is taken here
to be constant with respect to $r$, making
the integrals defined in (\ref{energy_a_definitions})
analytically tractable.
None the less the expressions, which have been verified with the aid of Mathematica 5.0,
are very long and are not displayed here.
Only their essential features, in particular
their dependence upon $\alpha$ and $r_{\bf max}$, are addressed here.
\par
With $A = e^{i\pi/4}$ we find that ${\cal E}_a$ has the functional
form
\beq
{\cal E}_a(r,t;\alpha)  = \frac{e^{-\alpha 2 T}}{r^{3/2}}\Upsilon(T,\cos{2T},\sin{2T};\alpha),
\eeq
where $\Upsilon$ is a well-defined analytical albeit complicated function of its arguments.
Thus,  ${\cal E}_a$  depends on time only through the variable $T=t/r^{3/2}$,
which is the time  measured in units of the local disc rotation period, at $r$,
divided by $2\pi$.
Different values
of $r$ merely change the overall amplitude of the response as governed by the coefficient
multiplying the function $\Upsilon$.  Otherwise, the evolution of ${\cal E}_a$ in
the disc is self similar and is always decaying at long times $T$.
Because of this we display below the evolution of
${\cal E}_a$ for different values of $\alpha$ at $r=1$ only.
\par
\begin{figure}
\begin{center}
\leavevmode \epsfysize=8.cm
\epsfbox{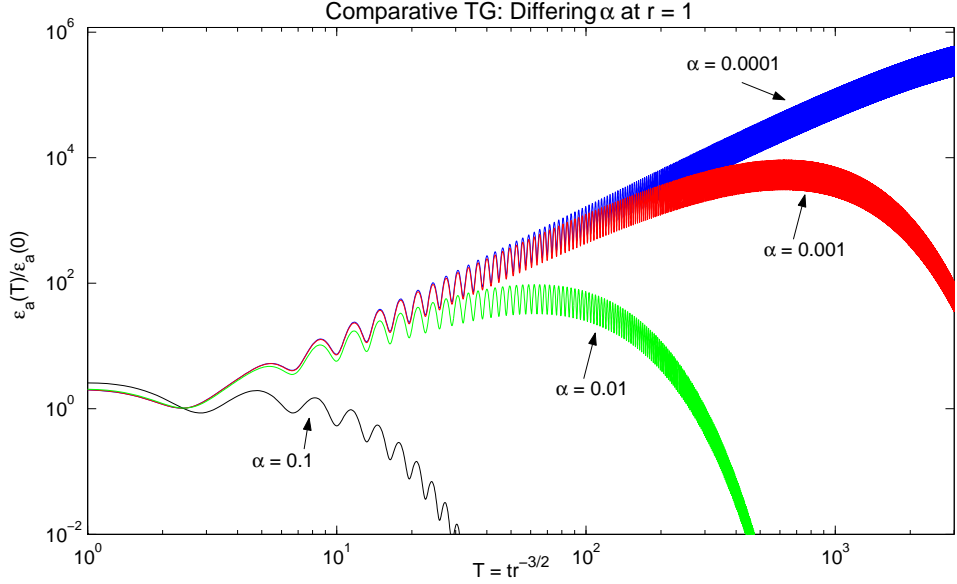}
\end{center}
\caption{{\small
The transient growth in the quantity ${\cal E}_a$ for the fundamental mode
at $r=1$, $n=3/2$ and where $A(r) = e^{i\pi/4}$, $dA/dr = 0$.  The four curves correspond to
$\alpha = 0.1, 0.01, 0.001, 0.0001$ and are presented
on a log-log plot.  All curves are ${\cal E}_a$ scaled to its corresponding
value at $T=0$.
The general rise time is proportional
to the inverse of $\alpha$ as well as the maximum amplitude achieved.
${\cal E}_a$ is also characterized by  oscillations with period $\pi$ that
sit atop the general transient trend.
}}
\label{transient_growth_plot1}
\end{figure}
Inspection of Figure \ref{transient_growth_plot1},
which
plots against $T$ the value of ${\cal E}_a(T)/ {\cal E}_a(0)$ for a range
of $\alpha$ values, uncovers a very important property of TG.
As $\alpha$ decreases the TG magnitude becomes more prominent
(for example, it is up to $\sim 3$ orders
of magnitude for $\alpha = 0.0025$) and the maximum occurs at
correspondingly later times. The time corresponding
to the maximum amplitude $T_{\rm max}$ is roughly $\sim 5/(8\alpha)$
in this\footnote{For the $k$-th overtone, $T_{\rm max}$ will be smaller
by a factor of $1/k^2$, see  (\ref{arbitrary_k_sigma})} case.
The rise in energy is modulated by oscillations which arise from
the fact that there are correlations between pairs of variables
which contribute to the TG of ${\cal E}_a(T)$ (see
Section \ref{discussion}).
These variables oscillate in a frequency defined by the imaginary
part of the eigenvalue (\ref{eigenvalue_inertioviscous_k0}) and since
products of pairs of variables terms depending on $2T$
appear, so that the period is half the orbital period
at the given radius.

\begin{figure}
\begin{center}
\leavevmode \epsfysize=10.cm
\epsfbox{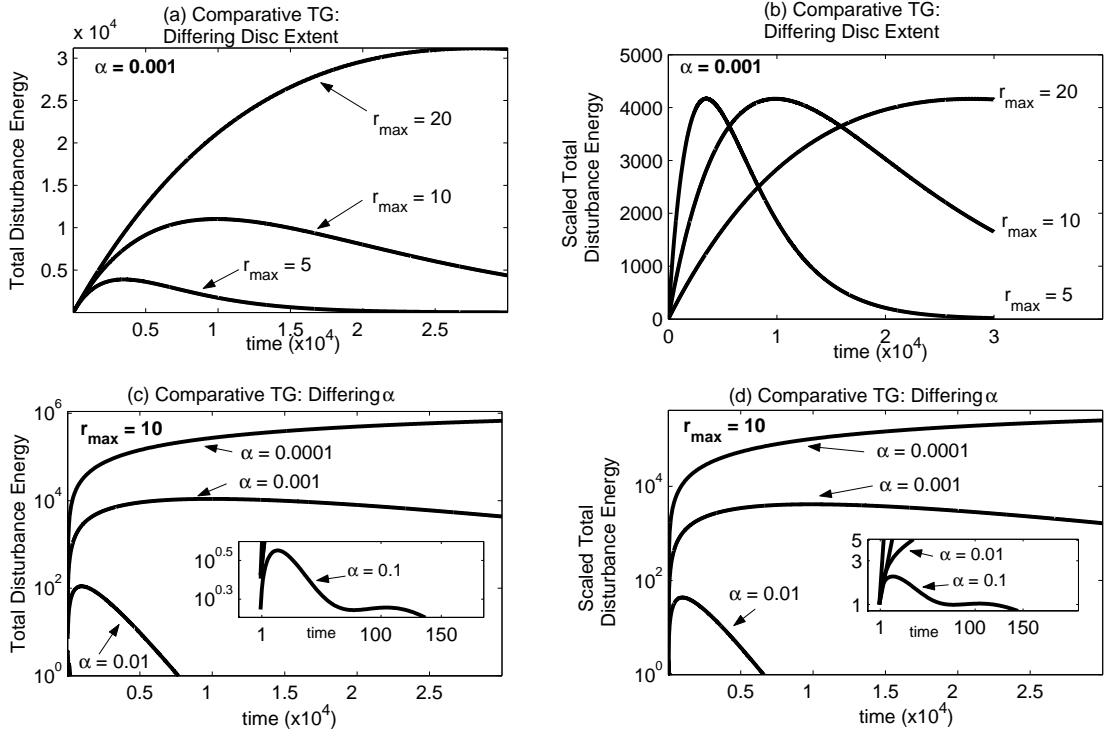}
\end{center}
\caption{{\small
Comparative TG behavior of the `total disturbance energy' ($E_a$)
and `scaled disturbance energy' ($E^{(s)}_a \equiv E_a(t)/E_a(0)$).
In all plots $n=3/2, r_{min} = 0.05$ and $A(r) = e^{i\pi/4}$
Panel (a) depicts $E_a$
for different values of $r_{\rm max}$ with $\alpha = 0.001$.
Larger values of $r_{\rm max}$ simply correspond to longer lived phenomena before
decay ultimately sets in at larger values of the disc radius
meaning that the total energies are larger.  In panel (b) $E^{(s)}_a$ is shown.
$E^{(s)}_a$ achieves the same overall maximum amplitude before decay.
Similarly panels (c) and (d)
depict $E_a$ and $E^{(s)}_a$ for $\alpha = 0.1,0.01,0.001,0.0001$ at $r_{max} = 10$.
The trend for the $E_a$ to grow with decreasing $\alpha$ is evident.  In (d) we see
that this growth, in terms of scaled energies,
can be dramatic - by nearly three orders of magnitude.}}
\label{transient_growth_plot3}
\end{figure}

In Figure \ref{transient_growth_plot3} we plot the behavior of $E_a$ for
assorted values of $\alpha$
and $r_{\rm max}$.  In all the cases displayed we have assumed that the inner radius,
$r_{\rm min}$, is $0.05$.   For given $r_{\rm max}$, smaller values of $\alpha$
correspond to more dramatic instances of TG.  The peak magnitudes
are proportional to the inverse of $\alpha$ as
is also the time when the peak occurs - a trend which is similar to that observed in
the quantity ${\cal E}_a$.
It is very important to keep in mind that
$A(r)$ affects the spatial details but not the global time behaviour. Finally,
we point out that the occurrence of the peak energy amplitude
is delayed for larger values of $r_{\rm max}$, at given values of $\alpha$, although
the value of the energy at the peak time remains roughly the same.

To get a sense for the expected qualitative {\em spatio}-temporal behavior of such
transiently growing solutions we have computed
the results for a case in which $A(r)$ is not constant (and thus the possibility
to evaluate the relevant integrals analytically can not be expected in general).
Assume that the initial radial velocity amplitude has a Gaussian form
centered around some fixed radius,
\beq
A(r) = e^{\imath \pi/4}e^{-\frac{(r-r_0)^2}{\Delta_f}}.
\eeq
In our particular case we take the full width at half maximum  given by $\Delta_f = 2.5$
and the initial center of the disturbance $r_0 = 3$.
Other parameters are $n=3/2$ and $\alpha = 0.0025$.
As pointed out above, with this form for $A$ we loose the ability to find
an analytic solution and consequently resort to numerical
evaluation.
In Figure \ref{DE_alphap0025} we show, in a contour plot in the disc meridional
plane, the kinetic energy (per unit volume) contained in the vertical
velocity disturbance $v_p$, that is, $ \rho_0 v_p^2/2$.
The transient growth and decay of the disturbance is shown in the
time sequence of figures displaying the spatial structure of the disturbance.
We also depict how the disc surface moves in response to the imposed
perturbation by solving the equation of motion for the boundary
at this order. In the Figure time is quoted in
terms of rotation times of the disc as measured at the radius $r=1$.
The two dramatic things to note is that,
(a) there maximal rise in the disturbance energy is of almost three orders of magnitude
and (b) the physical amplitude of fluctuations in the disc becomes quite large.
Associated with the
last effect is the emergence, as time moves on (by $t \sim 100$),
of highly crenellated patterns on the disc surface.
This crenellation is a result
of the epicyclic frequencies causing the amplitude profile to wind-up on itself.  Eventually, the
winding is so strong that viscous effects take over and wipe the perturbations away by a few thousand
rotation times.
\begin{figure}
\begin{center}
\leavevmode \epsfysize=12.cm
\end{center}
\caption{{\small
Transient growth in the disturbance kinetic energy density in a disc where
$n=3/2, \alpha = 0.0025$ and $A(r) = e^{i\pi/4}e^{\frac{(r-r_0)^2}{\Delta_f}}$,
with, in this example, $r_0 = 3$ and the full width-half-maximum $\Delta_f = 2.5$.
Time is quoted in terms of disc rotation times at the radius $r=1$ so that $T=t$.
The magnitude of of the disturbance energy density is indicated by the
various colours (whose meaning is indicated on the right of each panel)
on a cut through the disc's meridional plane.
A significant TG in the energy is observed (almost three orders of magnitude).
Strong crenellation patterns appear in the fluctuating boundary of the disc
before ultimate viscous decay takes over and wipes the pattern out.
}}
\label{DE_alphap0025}
\end{figure}
\subsection{The $\alpha \rightarrow 0$ limit: Algebraic growth}
The limit of zero viscosity, i.e. $\alpha \rightarrow 0$, results in
algebraic growth.
We find that for $n=3/2$, an arbitrary initial amplitude $A(r)$ and
the unapproximated disc height $h(r)$, the vertical velocity fluctuation $v_p'$ simplifies to
\beqa
\lim_{\alpha\rightarrow 0} v_p' &=&
e^{\imath T}\frac{Ah}{r}
\left(\frac{6}{7}\frac{r}{h}\frac{dh}{dr}-\frac{1711}{1470}
+\frac{5}{21}\frac{r}{A}\frac{dA}{dr}
-T\frac{5\imath}{14}\right)\zeta
 \nonumber \\
& & + e^{\imath T}\frac{Ah}{r}
\left(
\frac{8}{21}\frac{r}{h}\frac{dh}{dr}
+\frac{110}{147}
-\frac{4}{21}\frac{r}{A}\frac{dA}{dr}
+T\frac{2\imath}{7}\right)\zeta^3
\eeqa
The density follows in similar suit.  Notice that though the viscous effects vanish,
the oscillation frequency associated with the local epicyclic vibrations remain.
This solution appears as algebraically growing oscillations.
This result is consistent with the more general (than that with the
polytropic assumption) analysis made for discs with the fluctuations being
adiabatic (Umurhan \& Shaviv, 2005).\par
It is also important to note that the limiting process of $\alpha\rightarrow 0$
is one that cannot be taken without caution and care.  This is because
this is a singular limit - substituting simply $\alpha=0$
gives rise to a singularity in the expression
for the steady state height $h(r)$.  Inspection of the solution
to $h(r)$ as appearing in (\ref{disc_height}) shows that $h$
diverges as $\alpha \to 0$.  However, one could formally take this limit
holding the ratio $\dot M/\alpha$ fixed (cf. \ref{disc_height} and
the definition of $\Lambda$ in \ref{eq:for:hr}), namely
\begin{equation}
0 < \lim_{\alpha \rightarrow 0}   {{\dot M} \over \alpha} <\infty ,
\end{equation}
because the vanishing of the viscosity coefficient gives rise to a
vanishing mass transfer rate.
\section{Summary and Discussion}
\label{discussion}
This paper is a contribution to the ongoing investigation of the
purely hydrodynamic response of accretion discs to small perturbations.
Almost all of the recent studies, in which the phenomenon of transient
growth has been found, were based on local (using the `shearing box'
approximation) analysis (see references cited in Section \ref{introduction}).
While linear studies of this sort can be usually
done analytically, pursuing the perturbation development into the
non-linear regime calls for a fully numerical treatment.
It appears that the existing numerical studies
have not, as yet, achieved sufficient spatial resolution
(or were not fully three-dimensional) to be able
to definitely determine what is the ultimate dynamical fate of
the transiently grown (sometimes by several orders of magnitude) disturbances
(see, Yecko, 2004; Umurhan \& Regev, 2004;
Ashfordi {\em et} al., 2005; Mukhopadhyay {\em et} al., 2005).

In an endeavor to investigate disturbances which are not limited to
the shearing box (on which, typically, periodic boundary conditions are
imposed), but still be able to treat the problem analytically
or semi-analytically, we have resorted to asymptotic methods.
Adopting such a method, previously applied (in R and KK) to steady
thin (i.e. $\epsilon \ll 1$)
axisymmetric viscous discs  and extending it
to include time-dependence, we are able to show that transiently growing
disturbances naturally emerge on the global scale as well.
Moreover, we have found analytical expressions that help in
understanding the underlying physics giving
rise to the transient growth (see below).
In achieving this goal a number of simplifying assumptions
have been made. The unperturbed disc and the disturbances were assumed to be
polytropic and axisymmetric and they were treated in a ring
whose inner (outer) limit was significantly distant from the
inner (outer) physical disc boundary.
In addition, we chose to use the simplest initial
conditions so as to excite the most fundamental modes, which we
could readily analyze. As pointed out before these assumptions have
largely been made to allow for analytical treatment in the
framework of which
we have derived a set of dynamic equations for successive
orders of $\epsilon$. We stopped at the $\epsilon^2$ order,
but the expansion can be carried out to higher order terms.
The asymptotic expansion so defined brings out
the salient features which are dictated by a geometrically
thin cylindrical structure. Indeed, we were able to identify
two kinds of modes and obtained the analytical structure
of some simple initial disturbances.
To $\order{\epsilon^2}$
we have determined an analytical solution to the dynamic
problem and discovered
that this set of equations manifests strong TG,
irrespective of the polytropic index $n$.  
As we have noted,
the severity of the TG is inversely
proportional to $\alpha$ so that as the viscosity gets small,
the TG becomes more dramatic.
It seems then that TG is an intrinsic property of global
axisymmetric disturbances of polytropic discs and,
given the results of previous works done in the local
`shearing box' approximation as well as the two-dimensional
(in the disc plane) global case (Ioannaou \& Kakouris, 2002),
we may reasonably expect it to be a property of all
accretion discs.

In spite of the simplifying assumptions we have implemented,
and whose relaxation {\em must} be addressed in the future (see below),
it is important to remark that the asymptotic expansion is
not equivalent to the customary (in linear stability analysis)
linearization procedure of infinitesimal
perturbations imposed on a global steady state.
Although, to the orders to which solutions have
been determined the governing equations are linear,
this procedure is intrinsically
a finite-amplitude expansion (see Umurhan \& Shaviv, 2005).
Indeed, technically speaking, at each $\epsilon^n$ order, the equations are
linear with inhomogeneous terms that result from the $\epsilon^{n-1}$ order.
Yet the aggregate solution, and the region of its validity, put it in the
finite-amplitude realm.

We turn now to the investigation of the physical sources for the TG that we
have found in what we have called `general driven acoustic modes'. To this end
we consider the following energy density function
\beq
{\varepsilon}_a(r,z,t) \equiv \sfrac{1}{2}\rho_0{v'_{2}}^2 +
\sfrac{1}{2}\frac{c_{s0}^2}{\rho_0}{\rho'_{2}}^2.
\label{def:varepsilon_a}
\eeq
The first term in this expression represents the kinetic energy contained
in the time dependent part of vertical motion
while the second term represents a sort of thermal energy in the form of an acoustic disturbance
(Rayleigh, 1877). We have already considered integrals of
a similar function, but only for the particular solution $v_p'$ (see Section \ref{TG}).
Here we shall be interested in the temporal behaviour of the volume
integral of ${\varepsilon}_a$ over the entire domain of interest
$r_{\rm in}\le r \le r_{\rm out}$; $|z|\leq h$.

Using equations (\ref{v_mom_eq}-\ref{continuity_eq}) to substitute
for the relevant time-derivatives and converting the resulting integrals
to surface integrals whenever possible in the usual way, we get
\beq
\frac{d}{dt} \int_{\bf V}{\varepsilon_{_a} dV } =
-\int_{\bf S}{\left(\rho'_{2} v_{2}'c_{s0}^2 \right) dS}
+\sfrac{4}{3}\int_{\bf S}{\left(\eta v'_{2}\frac{\partial v'_{2}}{\partial z} \right) dS}
-\sfrac{4}{3}\int_{\bf V}{\eta \left(\frac{\partial v'_{2}}{\partial z}\right)^2 dS}
+T_E,
\label{acoustic_energy_integral}
\eeq
where $T_E$ represents volume integrals over the terms that are responsible for
transient energy growth and its explicit form is
\beq
T_E =
\int_{\bf V}{\left\{
\frac{v'_{2}}{r}\frac{\partial}{\partial r}\left(\eta  r \frac{\partial u'_{1}}{\partial z}\right)
-\frac{2}{3}v'_{2}\frac{\partial}{\partial z}\left[\frac{\eta }{r}\frac{\partial (ru'_{2})}{\partial r}\right]
-\frac{\rho'_{2}}{\rho_0}\frac{c_{s0}^2}{r}\frac{\partial( r\rho_0 u'_{1})}{\partial r}
\right\}
dV}.
\label{def:transient_expression}
\eeq
The first term on the RHS of (\ref{acoustic_energy_integral}) represents the
mechanical work done on the disc by the
bounding surface $S$.  But, given that (a)
the pressure of the moving boundary is zero and (b) the quantities
in the integrand are odd functions of $z$, this surface integral vanishes.
The second term  represents
the work performed on the surface by the viscous stresses
associated with gradients of the vertical
velocities.  In a similar fashion the surface integral of this term
also vanishes.
Its general nature is not immediately obvious, but given the solutions
we have obtained in this paper
we always find that it vanishes, mainly because of the fact that $\eta$ tends
to zero (when approaching the disc surface) sufficiently fast.
The third term is the volume integrated loss of energy due to
viscous dissipation, i.e. the {\em viscous losses}.
Since the integrand here is positive definite, this term is
always a sink of energy and is the ultimate cause of decay in all the perturbations.
Thus we concisely and transparently write the last relation as,
\[
\frac{d}{dt} \int_{\bf V}{\varepsilon_{_a} dV } =
T_E - {\rm [viscous \ losses].}
\]

Consider now the structure of the expression for $T_E$ in more detail.
The first two terms in (\ref{def:transient_expression})
are explicitly viscosity dependent, while the
third is not, although it may depend on $\alpha$
implicitly, via the $\alpha$ dependence of the disturbances.
The first two terms consist of the cross-correlations
between the vertical and horizontal viscous shear disturbances
on one hand and the vertical velocity disturbance on the other
hand.
Regarding the third term, we notice that it follows
from (\ref{relation_of_rhoprime_csprime}) that
${\rho'c_{s0}^2}/{\rho_0}$ is proportional to the dynamical
disturbance of the sound speed,
i.e., ${c_{s2}^2}'$ (or, equivalently, of the temperature),
so that the expression in the
integrand may be rewritten as
\[
{c_{s2}^2}'\frac{1}{r}\frac{\partial (r\rho_0 u'_1)}{\partial r}.
\]
The volume integral of this expression thus depends
on the correlation between the dynamical fluctuations
in the radial mass flux and its radial derivative
with the disturbances in the sound speed.
Although we have chosen only
disturbances which do not contribute at all to the total horizontal
mass transfer rate (see the condition expressed in equation
\ref{eq:zeromass}), the correlation integral consisting of the
third term obviously need not vanish. In other words,
the integral condition (\ref{eq:zeromass}) on the radial velocity
dynamical disturbance in the radial velocity does not imply that the
contribution to the work done on the fluid, as given in the third term,
is zero.

To assess the relative contribution of the first two viscous terms to
above discussed third correlation integral,
we calculated the relevant ratio
and found that for the three values of $\alpha$ displayed
in Figure \ref{transient_growth_plot1}, this ratio
decreases monotonously with $\alpha$. For example,
for $\alpha = 0.001$, it is less than $0.1\%$ for all $T>0$.
Thus,  for $\alpha \rightarrow 0$ the TG
is totally dominated by the correlations embodied in the third term.
This means that the algebraic growth found in the inviscid limit
by Umurhan \& Shaviv (2005) must be driven by this term.

We conclude with some remarks on possible improvements to the asymptotic analysis
of the sort done here and prospects for the future.
Asymptotic expansions, when viable, are often very robust and provide
a good approximation to the solution when truncation to only few leading
terms is done. Obviously, when a term in the series becomes {\em very} large
it may `break its order', that is, become of the order of a previous term and
as such make the expansion invalid in this region. In our expansions
successive terms ratios are of $\order{\epsilon^2}$, and thus the procedure's
validity should not severely be limited even up to a growth factor of
$\sim 100$ or so (in the velocity or density perturbations). Still,
the procedure, when carried to higher order, introduces corrections which are
technically non-linear.  Careful consideration must be undertaken in order to handle the
response at these higher orders.  This may entail treating the disturbance amplitudes for
the lower order solutions (like
$A(r)$ in $u_{1}'$) as {\em weakly non-linear} governed by a second `slow' time (e.g.
the amplitude is instead written as $A = A(r,\tau)$ where $\tau = \epsilon^2 t$) in a manner
analogous to the treatment of non-linear thick polytropes
(for example Balmforth \& Spiegel, 1996).

The approach used here may be generalized in a number of additional directions.
Allowing for non-axisymmetric perturbations, including the disc inner
and outer boundary in some kind of boundary layer analysis and relaxing
the polytropic assumption seem to be the most obvious generalizations.
In such future studies the implications of the presence of backflows in the
steady state should be examined as well. In any case, we have
found the presence of TG in the simplest cases. It is difficult to
imagine that it will be suppressed in the more general conditions
although the effect of radiative energy losses on TG must be
carefully examined.

The question concerning the ultimate fate of the transiently grown
perturbations and their ability to induce a sustained turbulent state
in the disc remains open. In this context it is worthwhile to notice
that since the TG decay times are of the order of hundreds of rotation periods,
it is conceivable that accretion discs, which are usually not isolated
systems, may experience recurrent external perturbations on such
time scales and in this way the dynamical activity may be sustained.

Extensive numerical calculations of
as realistic as possible accretion discs are however needed to decide
if TG {\em per se} may lead, through non-linear processes, to sustained turbulence.
Such extremely high-resolution global three-dimensional
calculations seem, however, to still be above the ability of the
present computer power and it may be advantageous to also exploit
sophisticated non-linear asymptotic methods to complement and guide them.

\appendix
\section{Disc flow solutions for arbitrary polytropic index}
We present here, for the sake of completeness, the steady-state solution,
which can be found in KK only for a particular value of the polytropic index.
\label{steady_state_KKsolutions}
\subsection{Radial velocity solution and the disc height as function of $r$}
One begins by combining (\ref{SS_epsilon2_ueqn}-\ref{SS_epsilon2_Omegaeqn}) into
a single equation for $u_1$.  After a little manipulation and upon using
(\ref{polyrelations}) we find:
\beq
\Omega_0^2\left\{\alpha^2\left[\frac{(h^2-z^2)}{2(n+1)}\frac{\partial^2}{\partial z^2}
-z\frac{\partial}{\partial z}\right]^2 + 1\right\}u_1 =
- \frac{2\Omega_0}{r^2\rho_0}\frac{\partial}{\partial r}\left(
\eta r^3\frac{\partial\Omega_0}{\partial r}
\right),
\label{u1_eqn}
\eeq
where $h(r)$ is the (still unknown) disc height.
The solution to (\ref{u1_eqn}) is
\beq
u_1(r,\zeta) = \frac{\alpha h^2}{r^{5/2}}
\left\{ \frac{(3n+1) [ 9(n+1)(1-\zeta^2)+8(2n+3)\alpha^2 ]}
{9(n+1)^2 + 4(2n+3)^2\alpha^2}
-\frac{2r}{h}\frac{dh}{dr} \right\}
\label{u1_solution}
\eeq
where $\zeta = z/h$.

We are now in a position to determine the disc height as
a function of the radius.  First,
we see that
vertically integrating the continuity equation
(\ref{SS_epsilon2_continuityeqn}), after multiplying
it by $2\pi r$, leads to a constraint whose physical
meaning is that the mass flowing through the disc is
independent on $r$.  In particular, we are left with
\beq
2\pi\int_{-h}^{h}dz{\frac{\partial}{\partial r}(r \rho_0 u_1)}
= 0,
\label{integral_constraint}
\eeq
because it is supposed a priori that the quantity $\rho_0 v_2$ indeed vanishes
as $z \rightarrow \pm h(r)$.
 The satisfaction of
(\ref{integral_constraint}) is equivalent to saying that
\[
\dot M_1 = {\rm constant} =
-2\pi\int_{-h}^{h}{dz (r\rho_0  u_1)},
\]
where $\dot M_1$ is the mass accretion rate, which is
one of the fundamental free parameters in this system\footnote
{Note that the minus sign in the above integral statement reflects
the convention that a positive mass accretion rate $\dot M_1
> 0$ implies physically that the
integrated mass flux must be a negative in order for
the material to flow {\em inward}.}.

The next step is the evaluation of the integral in the above equation
using (\ref{u1_solution}). It leads
to the following first order non-linear ODE for $h(r)$
\beq
\frac{h^{2n+3}}{r^{3n+3/2}}
\left(1-\frac{2n+3}{3n+1}\cdot\frac{r}{h}\frac{dh}{dr}
\right)=\Lambda
\equiv \frac{\dot M_1}{\alpha}\frac{\Gamma(n+5/2)}{\Gamma(n+1)}
\cdot \frac{(2n+2)^n}{\pi^{3/2}(3n+1)}.
\label{eq:for:hr}
\eeq
Though the above is non-linear, its solution is remarkably easy to
write down and reads
\beq
h(r) =
(2\Lambda)^{\frac{1}{2n+3}}r^{\frac{3n+1}{2n+3}}
\left(\sqrt r - \sqrt r_*\right)^{\frac{1}{2n+3}}.
\label{disc_height}
\eeq
This solution is shaped so that $r_*$ is the radius at which the disc
experiences zero torque.  At this radius our solution is singular
(the thickness of the disc becomes zero and the radial flow infinite).
For this reason
we suppose that $r_*$ is always significantly smaller than
the fiducial radius $r=1$ and never
consider these solutions near the zero torque point\footnote{None the less, it is
clear that the infinities appearing near $r_*$
indicate that these solutions breakdown in that vicinity and that a boundary-layer type
analysis will be required to cope with the failure.  By restricting ourselves to
be sufficiently far away from this pathological point we may proceed
with the subsequent analysis.}
Additionally, at the fiducial radius the height of the disc is given as
\beq
h_1 = h(1) =(2\Lambda)^{\frac{1}{2n+3}}
\left(1- \sqrt r_*\right)^{\frac{1}{2n+3}}.
\label{height_at_1}
\eeq
If $h_1$ is chosen to be some value, then there is a unique ratio of $\dot M/\alpha$ which
satisfies this constraint.

Some things to note with regard to the disc thickness function:  when $n=3/2$ the height of the disc
varies linearly with radius  for $r \gg r_*$;  when $n>3/2$ the structure of the
disc is {\em flared}, meaning to say $\frac{d}{dr}\frac{h}{r} > 0$ as
$r$ gets large.  The
overall `thickness' of the thin disc, in which $h_1$ is its measure,
is related to the accretion rate according to
\[
h_1\sim \dot M_1^{\frac{1}{2n+3}}.
\]
In the case where $n=3/2$, the scale of the disc's thickness
is proportional $\dot M_1^{1/6}$ which is consistent with the results
obtained by KK.

In conclusion of this part of the Appendix
we derive the general condition for the occurrence of backflow (see KK, RG).
In other words, we find the highest value of $\alpha$ for which
the radial velocity switches sign at some point in the midplane of the disc?
This may be determined by setting to zero $\zeta$ in (\ref{u1_solution}) and solving
the result for $\alpha$.  After making
use of (\ref{disc_height}) we find that $\alpha$ and the position $r$
at which the radial velocity in the midplane is zero are related
through the equation
\beq
\sfrac{2}{3}\alpha -
\sqrt{\frac{2n(n+1)}{(2n+3)^2}\left(1-\frac{3n+1}{2n}\sqrt{\frac{r_*}{r}}
\right)
} = 0.
\label{stagnation1}
\eeq
There exists thus a critical value of $\alpha = \alpha_\infty$, that is,
the one obtained for $r\to \infty$ in (\ref{stagnation1}) and it is given by
\[
\alpha_\infty \equiv
\frac{3}{2}\sqrt{\frac{2n(n+1)}{(2n+3)^2}}.
\]
Its significance is such that for values of $\alpha > \alpha_\infty$
there is no possibility of backflow in the disc.
For values of $0 < \alpha < \alpha_\infty$ there are places in the disc, which includes the midplane,
in which the flow is directed outwards.  This flow transition, or stagnation point,
bifurcates out of $r = \infty$ at
the value of $\alpha=\alpha_\infty$.
For an $n=3/2$ polytrope, $\alpha_\infty \sim 0.685$, which is
the value predicted by KK.
For values of $\alpha < \alpha_\infty$ the stagnation point, $r_{\rm stag}$ occurs at
\beq
\frac{r_{\rm stag}}{r_*} = \frac{81}{4}\frac{(n+1)^2(3n+1)^2}{[9n(n+1)-2(2n+3)^2\alpha^2]^2},
\eeq
which, in other words, means to say that
for a given value of $\alpha<\alpha_\infty$, backflow occurs only for values of $r>r_{stag}$.
Finally, there is also a minimum radius $r_{\rm stag}^{\rm min}$, below which there
is never backflow.  This
occurs in the limit where $\alpha\rightarrow 0$ and is given by,
\beq
\frac{r_{\rm stag}^{min}}{r_*} = \left(\frac{3n+1}{2n}\right)^2.
\eeq
In other words it says that $r_{\rm stag}$ appears in the $\alpha \rightarrow 0$ limit at the value
$r_{\rm stag}^{\rm min}$.  In the case $n = 3/2$, this absolute minimum stagnation point appears
at $(121/36)r_*$ (see KK).
\subsection{$\Omega_2, v_2$ and $\rho_2$}
The general solution to $\Omega_2$ comes from inserting $u_1$ into
(\ref{SS_epsilon2_Omegaeqn}) and solving.  We find that,
\beq
\Omega_2 =
\frac{h^2}{4r^{7/2}}\left[
(1-\zeta^2)\frac{2r}{h}\frac{dh}{dr}-3\left(1-
\frac{4(3n+1)(1-(2n+3)\zeta^2)\alpha^2}
{9(n+1)^2 + 4(2n + 3)^2 \alpha^2}\right)
\right].
\label{SS_Omega_2_generaln}
\eeq
Notice that the correction $\Omega_2$ to the rotational speeds is
non-zero in the limit of $\alpha \rightarrow 0$.
This merely represents the fact that at this order the
pressure contributes something to the overall
equilibrium of the system and this contribution makes the flow
slightly {\em sub}-Keplerian.  Notice
that unlike $\Omega_0$, the solution for $\Omega_2$ shows that
the angular velocity
is no longer constant on cylinders since
$\partial\Omega_2/\partial \zeta \neq 0$ in general.

Using (\ref{polyrelations}) for $\rho_0$, together with (\ref{eq:for:hr}),
one may straightforwardly, though cumbersomely,
integrate (\ref{SS_epsilon2_continuityeqn}) to yield a solution for
the vertical velocity. The result is
\beq
v_2 =
\frac{\alpha\zeta h^3}{2r^{7/2}}\left[
\left(\frac{2r}{h}\frac{dh}{dr}\right)^2 -
\frac{8(2n+3)(3n+1)\alpha^2}{9(n+1)^2 + 4(2n+3)^2\alpha^2}
\left(\frac{2r}{h}\frac{dh}{dr}\right)
-
27\frac{(n+1)(2n+1)(3n+1)(1-\zeta^2)}{(2n+3)(9(n+1)^2 + 4(2n+3)^2\alpha^2)}
\right].
\eeq
The expression for the density is quite lengthly and we only present it for the case
$n=3/2$.  The integration constants in deriving
$\rho_2$ were chosen so that $c_{s0}^2\rho_2/\rho_0 \rightarrow 0$ as $\zeta \rightarrow 0$,
\beq
\rho_2 =
-\frac{(1-\zeta^2)^{3/2} h^5}{360\sqrt 5 r^{13/2}}
\left[3\frac{
675(1+\zeta^2) + 16\alpha^2(460+528\alpha^2-189\zeta^2)}{25+64\alpha^2}
-16\alpha^2\frac{1665+2432\alpha^2}{25+64\alpha^2}\cdot\frac{r}{h}\frac{dh}{dr}
+ 192\alpha^2
\left(\frac{r}{h}\frac{dh}{dr}\right)^2
\right].
\eeq

\section{Detailed calculations for $\order{\epsilon^2}$ dynamics}
\subsection{Vertical acoustics}
\label{homo_acoustics_details}
We begin this analysis with an exploration of the
dynamics arising from the solution
to the homogeneous part of equation (\ref{combo_v_eq}),
\beq
{\cal L}v_h' = 0.
\label{homo_acoustic_eqn}
\eeq
Recalling the arguments pointed out in the text, in this circumstance
the vertical velocity and density fluctuations are completely
decoupled from the horizontal (radial and azimuthal) motions.
It is in this context that we consider the simplest vertical motions
possible,
Given that there are no horizontal velocities induced by
these motions, the boundary condition
(\ref{lagrangian_pressure_BC})  reduces to
\begin{equation}
\left. {P_0{{\partial v_h'} \over {\partial z}}} \right|_{z=h}=0
\end{equation}
Assuming the solution Ansatz (\ref{homo_acoustics_ansatz}) we find that
(\ref{homo_acoustic_eqn})  reduces to
\beq
-\frac{1}{n}\left(
1 + \frac{8}{9}\frac{\tilde\sigma\alpha n}{n+1}
\right)
\left[
(1-\zeta^2)\frac{\partial^2}{\partial \zeta^2} -2(n+1)\zeta\frac{\partial}{\partial \zeta}
-2n\frac{\tilde\sigma^2 + 1}{1 + \frac{8}{9}\frac{\tilde\sigma\alpha n}{n+1}}\right]\hat v_h' = 0,
\label{homo_acoustic_eqn_revealed}
\eeq
where for convenience we have introduced the height scaled variable $\zeta \equiv z/h$.

It is worth reviewing the behavior of possible solutions to the operator
(\ref{homo_acoustic_eqn_revealed}) near the singular points $\zeta = \pm1$.
In particular,
if we assume that the solution functionally behaves according to the form
\[
\hat v_h' \sim (1-\zeta^2)^\lambda,
\]
then an indicial equation analysis (Morse \& Feshbach, 1953) shows that
to leading order there are two values of $\lambda$ (corresponding to two linearly
independent solutions) which are $0$ and $-n$.  The former choice tells us that
there is, in fact, a solution described by a regular series expansion near $\pm 1$.
As for the latter solution, since it predicts that $\hat v_h'$ diverges like $\zeta^{-n}$
the product $P_0 \partial \hat v_h'/\partial z$ none the less remains constant.  However
this solution must still be rejected because it would cause the zero Lagrangian pressure
condition on the surface to be violated, i.e.
(\ref{order_epsilon2_pressurecondition})
would not be satisfied.

As it turns out, the solution to (\ref{homo_acoustic_eqn_revealed}) is composed of
the associated Legendre  polynomials $P$ and $Q$ (Abramowitz \& Stegun, 1972)
as follows,
\beqa
\hat v_h' &=& \frac{\tilde A(r)}{\left(1- \zeta  \right)^{n/2}}\cdot
\left[
P_{_\Delta}^n\left( \zeta  \right)
Q_{_\Delta}^n\left(0\right)
-
Q_{_\Delta}^n\left( \zeta  \right)
P_{_\Delta}^n\left(0\right)
\right], \nonumber \\
{\rm with}\nonumber\\
\Delta(\tilde\sigma,n,\alpha) &\equiv& \frac{1}{2}\left\{
-1 + \left[1 + \left((2n+1)^2-\frac{8n(n+1)(1+\tilde\sigma^2)}{n+1+\sfrac{8}{9}n\alpha\tilde\sigma)}
\right)\right]^{1/2}
\right\},
\eeqa
and this solution satisfies the condition (\ref{order_epsilon2_pressurecondition}).
The $z$ independent constant $\tilde A(r)$ is entirely arbitrary and is
determined by the initial conditions.
In order to satisfy the boundary condition at $z=h$ a somewhat complicated
expression emerges for the eigenvalue condition.  However, it turns out
that for special cases of the polytropic index
the quantization condition for these modes may be easily written down,
namely,
for $n$ a positive integer or half odd integer it is required that,
\beq
\Delta = 2k + n + 1,
\eeq
where $k$ is any integer $k\ge 0$.  In this sense $k$
labels the acoustic overtone of the mode with $k=0$ being the fundamental.
Given the definition of $\Delta$ above, for every $k$
there are two values of $\tilde\sigma$ which are always decaying.
In considering all fundamental modes, which we label with
$\tilde\sigma_{_F}$, we find that,
\beqa
\tilde\sigma_{_F} &=& -\frac{4}{9}\alpha \pm
\left(\frac{16}{81}\alpha^2 -
\frac{2n+1}{n}
\right)^{1/2} \nonumber \\
&\longrightarrow & -\frac{4}{9}\alpha \pm i \left|
\left(\frac{16}{81}\alpha^2 -
\frac{2n+1}{n}
\right) \right|^{1/2}, \ {\rm for} \ \alpha < \alpha_{\rm osc}.
\eeqa
It is interesting to note that for $\alpha < \alpha_{\rm osc}$, given by,
\[
\alpha_{\rm osc} = \frac{9}{4}\sqrt{\frac{2n+1}{n}},
\]
the modes are complex conjugate
pairs of decaying oscillations.  For  $\alpha>  \alpha_{osc}$
 the modes are strictly exponentially decaying.
For example, for $n=3/2$,  $\alpha_{osc} = (9/4)\cdot \sqrt{8/3}
\approx 4$.
Furthermore, in the zero viscosity limit, the oscillation frequency
is given by
\[
\tilde\sigma_{_F}(\alpha \rightarrow 0) = \pm i\sqrt{\frac{2n+1}{n}}.
\]

\subsection{General Driven Acoustics}
\label{inertio_viscous_modes}

We now turn to solving the coupled equations (\ref{single_u_eqn}-\ref{combo_v_eq}).
Given the structure of these equations we must begin with developing a solution to
(\ref{single_u_eqn}) and then using this in (\ref{combo_v_eq}).
We suppose, as we did in the previous section, that the solution
to (\ref{single_u_eqn}), which is a homogeneous equation,
has the following form,
\beq
u'_1 = \hat u'(z,r)e^{\sigma T} + {\rm c.c.}
\label{uprime_solution}
\eeq
We write the eigenvalue as $\sigma$ in order to distinguish it from the
`vertical acoustics'
eigenvalue, $\tilde\sigma$, discussed in the previous section.
After a little manipulation we find that  (\ref{single_u_eqn}) turns into the
far simpler ODE
\beq
\Omega_0^2 \left\{ \left[ \sigma - \frac{\alpha}{2(n+1)}(1-\zeta^2)
\frac{\partial^2}{\partial \zeta^2}
+\alpha\zeta\frac{\partial}{\partial \zeta}\right]^2 + 1\right\}\hat u' = 0,
\label{Poperator}
\eeq
\par
Before presenting the Ansatz for the solution it is worthwhile to consider the
array of solution behavior of (\ref{Poperator}) near the singular points $\zeta = \pm 1$.
Assuming that near $\zeta =1$ the behavior is $\hat u' \sim (1-\zeta^2)^\beta$,
an indicial analysis (Morse \& Feshbach, 1953) reveals that there are four choices
for $\beta$, namely, $0,1,-n,1-n$.  The solutions with $\beta = -n,1-n$ are rejected on grounds
that they violate the boundary condition at $\zeta = \pm 1$
(they essentially diverge). In other words, since $\eta$
is proportional to $P_0$, then as one approaches the boundary, $\eta(\partial u'/\partial z), \
\eta r(\partial \Omega'/\partial r)
\neq 0$ there for these two values of $\beta$. It would
then mean that the stress conditions (\ref{order_epsilon2_stresscondition})
would be violated there and so these solutions are ignored.
The remaining two possibilities suggest that a regular truncated series expansion is
a satisfactory means of expressing the solution to $u'$.

Thus, for varicose modes we assume the Ansatz
\beq
\hat u' =
\hat u_{(k)} = \sum_{m=0}^{k+1}A_m(r) \zeta^{2(k+1)-2m},
\label{solution_form_u}
\eeq
where $k$ labels the eigenmode and is an integer greater than or equal to zero. Correspondingly,
the eigenvalue, $\sigma$, would be labelled by $k$ too.
Aside from a generalized characterization of the eigenvalues for arbitrary $k$
(see below),
{\em from here on we
restrict ourselves to considering only the fundamental mode, i.e. $k=0$}
and thus,
\beq
\hat u' = \hat u_{(0)} =
A(r) \zeta^2 + B(r),
\label{fundamental_uform}
\eeq
where the constants have been redefined for simplicity.
Generally speaking, insertion of the solution form
(\ref{solution_form_u}) into (\ref{Poperator})
requires one to recursively solve for the values of the coefficients $A_k$ in the usual
procedure called for when developing series solutions to ODE's.  The first of these equations
is always a relationship for the eigenvalue $\sigma$.  For example inserting (\ref{fundamental_uform})
into (\ref{Poperator}) gives
\[
A\left[\left(\sigma + \alpha\frac{2}{3}\frac{2n+3}{n+1}\right)^2 + 1\right]\zeta^2
+\sigma\left(\sigma B-A\frac{2}{3}\frac{\alpha}{n+1}\right)
-A\frac{2}{3}\frac{\alpha}{n+1}\left(\sigma + \frac{2}{3}\alpha\frac{2n+3}{n+1}\right) + B = 0,
\]
and by setting to zero the coefficients of each power of $\zeta$ one gets
\beq
\sigma = \sigma_\pm = \pm i - \alpha\frac{2n+3}{n+1},
\qquad
\frac{B(r)}{A(r)} = -\frac{1}{2n+3}.
\label{inertio_viscous_temporal_response}
\eeq
In other words, there are the two possible temporal responses
(denoted by $\sigma_{\pm}$),
to this type of disturbance, however, the coefficients are the
same for both.
We note that the eigenvalue form indicates
that these modes exhibit decaying
oscillations.  For general values of $k$, the frequency response obtained by inserting
(\ref{solution_form_u}) into (\ref{Poperator}), is given by
\beq
\sigma_k = \sigma(k) = \pm i  -  \frac{(k+1)(2n+3+2k)}{(n+1)}\frac{2}{3}\alpha ,
\label{arbitrary_k_sigma}
\eeq
showing that these modes become increasingly damped in proportion to $k^2$ in the large $k$ limit.
The general solution for an arbitrary disturbance
can thus be written as
\[
u'_1 =
\sum_{k=0}^{\infty}
\hat u_{(k)}(\zeta,r)e^{\sigma_k T} + {\rm c.c.}.
\]
We note also that the integrated radial flux of mass
{\em for the perturbation}
vanishes, namely that,
\beq
\int_{-1}^{1}\rho_0 r \hat u' d\zeta= 0.
\eeq
and this is generally true for any integer mode $k$.  The consequence
being, as will be seen below,  that the radial structure of the perturbation amplitude,
namely the quantity $A(r)$, remains  a freely chosen function, usually guided by some
initial condition.
\par
Now we turn to expanding  the particular solution to (\ref{combo_v_eq}) armed
with the solutions (\ref{uprime_solution}),(\ref{fundamental_uform})
and
(\ref{inertio_viscous_temporal_response}).
Inspection of the operator structures of ${\cal L},
{\cal F}$ and $ {\cal G}$ suggests
that the solution to the particular vertical velocity, $v_p'$, has the form
\beq
v_p' =
\sum_{k=0}^{\infty}
(\hat v_{(k)}(\zeta,r)e^{\sigma T}
+ \hat V_{(k)}(\zeta,r)T e^{\sigma T}) \ \ +  {\rm c.c.},
\eeq
in which
\beq
\hat v_{(k)} = \sum_{m=0}^{k+1}\Xi_{k,2m}(r) \zeta^{2(k+1)-2m +1},
\qquad
\hat V_{(k)} = \sum_{m=0}^{k+1}\Theta_{k,2m}(r) \zeta^{2(k+1)-2m +1}.
\label{solution_forms_v}
\eeq
We note that these solutions have terms which manifest TG
accompanied by oscillations
due to the form $te^{-\alpha t}e^{i\omega t}$ type of behavior.
Recalling that we are here only considering the fundamental mode,
the above solution form is written as
\beq
\hat v_{(0)} = \Xi_{00}(r) \zeta^3 + \Xi_{02}(r) \zeta,\qquad
\hat V_{(0)} = \Theta_{00}(r) \zeta^3 + \Theta_{02}(r) \zeta.
\label{solution_forms_v_fundamental}
\eeq
Using
(\ref{solution_forms_v_fundamental})
together with the solutions
(\ref{uprime_solution}),(\ref{fundamental_uform})
in (\ref{combo_v_eq}) and after some {\em lengthly} algebra (which was verified
using Mathematica 5.0), the coefficients
straightforwardly follow.  Unfortunately,
their expressions
are very long and we have omitted their presentation here.
This solution for $v_p'$ along with the solution for $u'_1$, when used in (\ref{continuity_eq})
immediately yields the solution to $\rho'_2$.  Its expansion is similar to that of
of $v_p'$, namely, we say that $\rho'_2 \sim \rho_p' + \rho_h'$, representing its particular
and homogeneous contributions.  Then it follows that
\beq
\rho_p' = \hat \rho'(\zeta,r)e^{\sigma T}
+ \hat R'(\zeta,r) T e^{\sigma T} \ \ +  {\rm c.c.},
\eeq
in which
\beq
\hat \rho' =
\hat \rho_{(k)} = \sum_{m=0}^{k+2}\Xi_{k,2m}^{(\rho)}(r) \zeta^{2(k+1)-2m},
\qquad
\hat R' =
\hat R_{(k)} = \sum_{m=0}^{k+2}\Theta_{k,2m}^{(\rho)}(r) \zeta^{2(k+1)-2m}.
\label{solution_forms_rho}
\eeq
The superscript $(\rho)$ appearing in the above expressions is meant as a label
to distinguish these functions
from the corresponding ones for the vertical velocity mode, $v_p'$, cf. (\ref{solution_forms_v}).
The lowest density fluctuation eigenmode is
\beq
\hat \rho_{(0)} = \Xi^{(\rho)}_{00}(r) \zeta^4 + \Xi^{(\rho)}_{02}(r) \zeta^2 + \Xi^{(\rho)}_{04}(r),\qquad
\hat R_{(0)} = \Theta^{(\rho)}_{00}(r) \zeta^4 + \Theta^{(\rho)}_{02}(r)\zeta^2 + \Theta^{(\rho)}_{04}(r).
\label{solution_forms_rho_fundamental}
\eeq

The solutions quoted in the text have this form.
We note also that the boundary conditions on the vertical velocity components, namely,
\beq
\left(1-\zeta^2\right)^{n+1}
\left[
\frac{1}{h(r)}\frac{\partial \hat v_p'}{\partial \zeta} +
\frac{1}{r}\frac{\partial r \hat u'_1}{\partial r} \right]
= 0 ,
\eeq
as $\zeta \rightarrow 1$,
is  satisfied by these  solutions.

\label{lastpage}
\end{document}